\documentclass[11pt]{article}
\usepackage{amssymb,amsmath}
\usepackage{epsfig}
\usepackage{palatino}

\newtheorem{theorem}{Theorem}

\newtheorem{definition}{Definition}

\newtheorem{claim}[theorem]{Claim}
\newtheorem{lemma}{Lemma}

\newcommand{\qedsymb}{\hfill{\rule{2mm}{2mm}}}
\newenvironment{proof}[1][]{\begin{trivlist}
\item[\hspace{\labelsep}{\bf\noindent Proof#1:\/}] }{\qedsymb\end{trivlist}}

\def\calC{{\cal C}}

\def\calH{{\cal H}}
\def\calL{{\cal L}}
\def\calM{{\cal M}}

\def\R{\mathbb{R}}

\def\C{\mathbb{C}}

\def\calS{{\cal{S}}}
\def\poly{{\rm poly}}
\def\mns{{\mbox{-}}}
\def\pls{{\mbox{+}}}
\def\cphase{{C_\phi}}

\def\ra{\rangle}

\newcommand{\be}{\begin{eqnarray}}
\newcommand{\ee}{\end{eqnarray}}
\newcommand{\smfrac}[2]{\mbox{$\frac{#1}{#2}$}}

\newcommand\defeq{\stackrel{def}{=}}

\newcommand\ket[1]{{ |{#1} \rangle }}
\newcommand\bra[1]{{ \langle {#1} | }}
\newcommand\ketbra[1]{{\ket{#1}\bra{#1}}}
\newcommand{\eff}{\mathrm{eff}}

\def\BQNP{{\sf{BQNP}}}
\def\QCMA{{\sf{QCMA}}}
\def\QMA{{\sf{QMA}}}
\def\NP{{\sf{NP}}}
\def\MA{{\sf{MA}}}
\def\P{{\sf{P}}}

\def\locHam{{\sc local Hamiltonian}}
\def\Yes{{\sc Yes}}
\def\No{{\sc No}}

\newcommand{\ignore}[1]{}

\newcommand{\eps}{\varepsilon}
\renewcommand{\epsilon}{\varepsilon}

\DeclareMathOperator{\Spec}{Spec}
\newcommand*{\pert}[1]{\widetilde{#1}}

\begin{document}

\title{\bf  The Complexity of the Local Hamiltonian Problem}

\author{
 Julia Kempe\\
CNRS \& LRI, Universit\'e de Paris-Sud, \\
91405 Orsay, France, and\\
UC Berkeley, Berkeley, CA94720
 \and
 Alexei Kitaev\\
Departments of Physics and Computer Science,\\
California Institute of Technology, \\
Pasadena, CA 91125
 \and
 Oded Regev\\
Department of Computer Science,\\
 Tel-Aviv University,\\
 Tel-Aviv 69978, Israel
}

\date{\today}

\maketitle

\begin{abstract}
The $k$-{\locHam} problem is a natural complete problem for the
complexity class $\QMA$, the quantum analog of $\NP$.  It is similar
in spirit to {\sc MAX-$k$-SAT}, which is $\NP$-complete for $k\geq
2$. It was known that the problem is $\QMA$-complete for any $k \geq
3$. On the other hand 1-{\locHam} is in {\P}, and hence not believed
to be $\QMA$-complete. The complexity of the 2-{\locHam} problem has
long been outstanding. Here we settle the question and show that it is
$\QMA$-complete. We provide two independent proofs; our first
proof uses only elementary linear algebra.
Our second proof uses a powerful technique for analyzing the sum of two Hamiltonians;
this technique is based on perturbation theory and we believe that it
might prove useful elsewhere.
Using our techniques we also show that adiabatic
computation with two-local interactions on qubits is equivalent to
standard quantum computation.
\end{abstract}

\section{Introduction}


Quantum complexity theory has emerged alongside the first efficient quantum algorithms in an attempt to formalize the
notion of an {\em efficient} algorithm. In analogy to classical complexity theory, several new quantum complexity
classes have appeared. A major challenge today consists in understanding their structure and the interrelation between
classical and quantum classes.

One of the most important classical complexity classes is $\NP$ - nondeterministic polynomial time. This class
comprises languages that can be {\em verified} in polynomial time by a deterministic verifier. The celebrated
Cook-Levin theorem (see, e.g., \cite{Papadimitriou:book}) shows that this class has {\em complete} problems. More
formally, it states that SAT is $\NP$-complete, i.e., it is in $\NP$ and any other language in $\NP$ can be reduced to
it with polynomial overhead. In SAT we are given a set of clauses (disjunctions) over $n$ variables and asked whether
there is an assignment that satisfies all clauses. One can consider the restriction of SAT in which each clause
consists of at most $k$ literals. This is known as the $k$-SAT problem. It is known that 3-SAT is still $\NP$-complete
while 2-SAT is in {\P}, i.e., has a polynomial time solution. We can also consider the MAX-$k$-SAT problem: here, given
a $k$-SAT formula and a number $m$ we are asked whether there exists an assignment that satisfies at least $m$ clauses.
It turns out that MAX-2-SAT is already $\NP$-complete; MAX-1-SAT is clearly in {\P}.

The class $\QMA$ is the quantum analogue of $\NP$ in a probabilistic setting, i.e., the class of all languages that can
be probabilistically verified by a quantum verifier in polynomial time (the name is derived from the classical class
$\MA$, which is the randomized analogue of $\NP$). This class, which is also called $\BQNP$, was first studied in
\cite{Knill:96d,Kitaev:book}; the name $\QMA$ was given to it by Watrous \cite{Watrous:00a}. Several problems in $\QMA$
have been identified \cite{Watrous:00a,Kitaev:book,Janzig:03a}. For a good introduction to the class $\QMA$, see the
book by Kitaev et al. \cite{Kitaev:book} and the paper by Watrous \cite{Watrous:00a}.

Kitaev, inspired by ideas due to Feynman,  defined the quantum analogue of the classical SAT problem, the {\sc local
Hamiltonian} problem \cite{Kitaev:book}.\footnote{For a good survey of the {\sc local Hamiltonian} problem see \cite{Aharonov:02a}.} An instance of $k$-{\locHam} can be viewed as a set of local constraints on
$n$ qubits, each involving at most $k$ of them. We are asked whether there is a state of the $n$ qubits such that the
expected number of violated constraints is either below a certain threshold or above another, with a promise
that one of the two cases holds and both thresholds are at least a constant apart. More formally, we are to determine
whether the {\em groundstate} energy of a given $k$-local Hamiltonian is below one threshold or above another.

Kitaev proved \cite{Kitaev:book} that the 5-{\locHam} problem is $\QMA$-complete. Later, Kempe and Regev showed that
already 3-{\locHam} is complete for $\QMA$ \cite{Kempe:03a}. In addition, it is easy to see that 1-{\locHam} is in
{\P}. The complexity of the 2-{\locHam} problem was left as an open question in
\cite{Aharonov:02a,Wocjan:03a,Kempe:03a,Bravyi:03a}. It is not hard to see that the $k$-{\locHam} problem contains the
{\sc MAX-k-SAT} problem as a special case.\footnote{The idea is to represent the $n$ variables by $n$ qubits and represent each clause by
a Hamiltonian. Each Hamiltonian is diagonal and acts on the $k$ variables that appear in its clause. It `penalizes' the
assignment that violates the clause by increasing its eigenvalue. Therefore, the lowest eigenvalue of the sum of the
Hamiltonians corresponds to the maximum number of clauses that can be satisfied simultaneously. } Using the known
$\NP$-completeness of MAX-2-SAT, we obtain that 2-{\locHam} is $\NP$-hard, i.e., any problem in $\NP$ can be reduced to
it with polynomial overhead. But is it also $\QMA$-complete? Or perhaps it lies in some intermediate class between
$\NP$ and $\QMA$? Some special cases of the problem were considered by Bravyi and Vyalyi \cite{Bravyi:03a}; however,
the question still remained open.

In this paper we settle the question of the complexity of 2-{\locHam} and show

\begin{theorem}\label{thm:main_thm}
The 2-{\locHam} problem is $\QMA$-complete.
\end{theorem}

In \cite{Kitaev:book} it was shown that the $k$-{\locHam} problem is in $\QMA$ for any constant $k$ (and in fact even
for $k=O(\log n)$ where $n$ is the total number of qubits). Hence, our task in this paper is to show that any problem
in $\QMA$ can be reduced to the 2-{\locHam} problem with a polynomial overhead. We give two self-contained proofs for
this.

Our first proof is based on a careful selection of gates in a quantum circuit and several
applications of a lemma called the {\em projection lemma}. The proof is quite involved; however, it uses
only elementary linear algebra and hence might appeal to some readers.

Our second proof is based on perturbation theory -- a collection of techniques that are used to analyze sums of
Hamiltonians. This proof is more mathematically involved. Nevertheless, it might give more intuition as to why the
2-{\locHam} problem is $\QMA$-complete. Unlike the first proof which shows how to represent any $\QMA$ circuit by a
2-local Hamiltonian, the second proof shows a reduction from the 3-{\locHam} problem (which is already known to be
$\QMA$-complete \cite{Kempe:03a}) to the 2-{\locHam} problem. To the best of our knowledge, this is the first reduction
{\em inside} $\QMA$ (i.e., not from the circuit problem). This proof involves what is known as {\em third order}
perturbation theory (interestingly, the projection lemma used in our first proof can be viewed as an instance of {\em
first order} perturbation theory). We are not aware of any similar application of perturbation theory in the literature
and we hope that our techniques will be useful elsewhere.

\paragraph{Adiabatic computation:} It has been shown in \cite{Aharonov:04a} that the model of adiabatic computation with $3$-local
interactions is equivalent to the standard model of quantum computation (i.e., the quantum circuit model).\footnote{
Interestingly, their proof uses ideas from the proof of $\QMA$-completeness of the {\sc local Hamiltonian} problem.} We
strengthen this result by showing that 2-local interactions suffice.\footnote{The main result of \cite{Aharonov:04a} is
that $2$-local adiabatic computation on {\em six-dimensional particles} is equivalent to standard quantum computation.
This result is incomparable to ours since their particles are set on a two-dimensional
grid and all two-local interactions are between closest
neighbors.} Namely, the model of adiabatic computation with $2$-local
interactions is equivalent to the standard model of quantum computation. We obtain this result by applying the
technique of perturbation theory, which we develop in the second proof of the main theorem.

\paragraph{Recent work:}
After a preliminary version of our paper has appeared \cite{Kempe:04a},
Oliveira and Terhal \cite{Terhal:05a} have generalized our results and have shown that the 2-{\locHam} problem remains
$\QMA$-complete even if the Hamiltonians are restricted to nearest neighbor interactions between
qubits on a $2$-dimensional grid. Similarly, they show that the model of adiabatic computation
with $2$-local Hamiltonians between nearest neighbor qubits on a $2$-dimensional grid is
equivalent to standard quantum computation.
Their proof applies the perturbation theory techniques
that we develop in this paper and introduces several novel ``perturbation gadgets" akin to our
three-qubit gadget in Section \ref{sec:gadget}.

\paragraph{Structure:}
We start by describing our notation and some basics in Section \ref{sec:QMA}. Our first proof is developed in
Sections \ref{sec:spacecutter}, \ref{sec:previous} and \ref{sec:2local}. The main tool in this proof, which we name the
projection lemma, appears in Section \ref{sec:spacecutter}. Using this lemma, we rederive in Section \ref{sec:previous}
some of the previously known results. Then we give the first proof of our main theorem in Section \ref{sec:2local}. In
Section \ref{sec:perturbation} we give the second proof of our main theorem. This proof does not require the projection
lemma and is in fact independent of the first proof. Hence, some readers might choose to skip Sections
\ref{sec:spacecutter}, \ref{sec:previous} and \ref{sec:2local} and go directly to Section \ref{sec:perturbation}. In
Section \ref{sec:adiabatic} we show how to use our techniques to prove that 2-local adiabatic computation is equivalent
to standard quantum computation. Some open questions are mentioned in Section \ref{sec:conclude}.

\section{Preliminaries}\label{sec:QMA}

$\QMA$ is naturally defined as a class of promise problems: A promise problem $L$ is a pair ($L_{yes},L_{no}$) of
disjoint sets of strings corresponding to {\Yes} and {\No} instances of the problem. The problem is to determine, given
a string $x \in L_{yes} \cup L_{no}$, whether $x\in L_{yes}$ or $x \in L_{no}$. Let ${\cal B}$ be the Hilbert space of
a qubit.
\begin{definition}[$\QMA$] \label{Def:QMA}
Fix $\eps=\eps(|x|)$ such that $\eps = 2^{-\Omega(|x|)} $. Then, a promise problem $L$ is in $\QMA$ if there exists a
quantum polynomial time verifier $V$ and a polynomial $p$  such that:
\begin{itemize}
 \item[-] $\forall x \in L_{yes} \quad \exists \ket{\xi} \in {\cal B}^{\otimes p(|x|)} \quad \Pr \left( V(\ket{x},\ket{\xi} )=1\right)\geq 1-\eps$
 \item[-] $\forall x \in L_{no} \quad \forall \ket{\xi} \in {\cal B}^{\otimes p(|x|)} \quad \Pr \left( V(\ket{x},\ket{\xi})=1\right)\leq \eps$
\end{itemize}
where $\Pr\left( V(\ket{x},\ket{\xi})=1\right)$ denotes the probability that $V$ outputs $1$ given $\ket{x}$ and
$\ket{\xi}$.
\end{definition}
We note that in the original definition $\eps$ was defined to be $2^{-\Omega(|x|)} \leq \eps \leq 1/3$. By using
amplification methods, it was shown in \cite{Kitaev:book} that for any choice of $\eps$ in this range the resulting
classes are equivalent. Hence our definition is equivalent to the original one. In a related result, Marriott and
Watrous \cite{Marriott:04a} showed that exponentially small $\eps$ can be achieved without amplification with a
polynomial overhead in the verifier's computation.

A natural choice for the quantum analogue of SAT is the {\locHam} problem. As we will see later, this problem is indeed
a complete problem for $\QMA$.
\begin{definition}
We say that an operator $H:{\cal B}^{\otimes n} \rightarrow {\cal B}^{\otimes n}$ on $n$ qubits is a $k$-local
Hamiltonian if $H$ is expressible as $H=\sum_{j=1}^r H_j$ where each term is a Hermitian operator acting on at most $k$
qubits.
\end{definition}
\begin{definition} \label{Def:localHam}
The (promise) problem $k$-{\locHam} is defined as follows. We are given a $k$-local Hamiltonian on $n$-qubits
$H=\sum_{j=1}^r H_j$ with $r=\poly(n)$. Each $H_j$ has a bounded operator norm $\| H_j \| \leq \poly(n)$ and its
entries are specified by $\poly(n)$ bits. In addition, we are given two constants $a$ and $b$ with $a<b$.
In {\Yes} instances, the smallest eigenvalue of $H$ is at most $a$. In {\No} instances, it is larger than $b$. We
should decide which one is the case.
\end{definition}


We will frequently refer to the lowest eigenvalue of some Hamiltonian $H$.
\begin{definition}
Let $\lambda(H)$ denote the lowest eigenvalue of the Hamiltonian $H$.
\end{definition}
Another important notion that will be used in this paper is that of a {\em restriction} of a Hamiltonian.
\begin{definition}
Let $H$ be a Hamiltonian and let $\Pi$ be a projection on some subspace $\calS$. Then we say that the Hamiltonian $\Pi
H \Pi$ on $\calS$ is the restriction of $H$ to $\calS$. We denote this restriction by $H|_\calS$.
\end{definition}

\section{Projection Lemma}\label{sec:spacecutter}

Our main technical tool is the {\em projection} lemma. This lemma (in a slightly different form) was already used in
\cite{Kempe:03a} and \cite{Aharonov:04a} but not as extensively as it is used in this paper (in fact, we apply it four
times in the first proof of our main theorem). The lemma allows us to successively {\em cut out} parts of the Hilbert space
by giving them a large {\em penalty}. More precisely, assume we work in some Hilbert space $\calH$ and let $H_1$ be
some Hamiltonian. For some subspace $\calS \subseteq \calH$, let $H_2$ be a Hamiltonian with the property that $\calS$
is an eigenspace of eigenvalue $0$ and $\calS^\perp$ has eigenvalues at least $J$ for some large $J \gg \|H_1\|$. In
other words, $H_2$ gives a very high penalty to states in $\calS^\perp$. Now consider the Hamiltonian $H=H_1+H_2$. The
projection lemma says that $\lambda(H)$, the lowest eigenvalue of $H$, is very close to $\lambda(H_1|_\calS)$, the
lowest eigenvalue of the restriction of $H_1$ to $\calS$. The intuitive reason for this is the following. By adding
$H_2$ we give a very high penalty to any vector that has even a small projection in the $\calS^\perp$ direction. Hence,
all eigenvectors with low eigenvalue (and in particular the one corresponding to $\lambda(H)$) have to lie very close
to $\calS$. From this it follows that these eigenvectors correspond to the eigenvectors of $H_1|_\calS$.

The strength of this lemma comes from the following fact. Even though $H_1$ and $H_2$ are local Hamiltonians,
$H_1|_\calS$ is not necessarily so. In other words, the projection lemma allows us to approximate a non-local
Hamiltonian by a local Hamiltonian.

\begin{lemma}\label{lem:leak}
Let $H=H_1+H_2$ be the sum of two Hamiltonians operating on some Hilbert space $\calH=\cal S+\cal S^\perp$. The
Hamiltonian $H_2$ is such that $\cal S$ is a zero eigenspace and the eigenvectors in $\cal S^\perp$ have eigenvalue at
least $J> 2\|H_1\|$. Then,
 $$\lambda(H_1|_\calS) - \frac{\|H_1\|^2}{J-2\|H_1\|} \le \lambda(H) \le \lambda(H_1|_\calS).$$
\end{lemma}

Notice that with, say, $J \ge 8\|H_1\|^2 + 2\|H_1\| = \poly(\|H_1\|)$ we have $\lambda(H_1|_\calS) - 1/8 \le \lambda(H)
\le \lambda(H_1|_\calS)$.

\begin{proof}
First, we show that $\lambda(H)\le \lambda(H_1|_\calS)$. Let $|\eta\ra \in \calS$ be the eigenvector of $H_1|_\calS$
corresponding to $\lambda(H_1|_\calS)$. Using $H_2|\eta \ra = 0$,
\begin{equation*}
\bra{\eta}H \ket{\eta}=\bra{\eta}H_1 \ket{\eta}+\bra{\eta}H_2 \ket{\eta}= \lambda(H_1|_\calS)
\end{equation*}
and hence $H$ must have an eigenvector of eigenvalue at most $\lambda(H_1|_\calS)$.

We now show the lower bound on $\lambda(H)$. We can write any unit vector $\ket{v} \in \calH$ as $\ket{v}=\alpha_1
\ket{v_1}+\alpha_2 \ket{v_2}$ where $\ket{v_1} \in \cal S$ and $\ket{v_2} \in \calS^\perp$ are two unit vectors,
$\alpha_1,\alpha_2 \in \R$, $\alpha_1,\alpha_2 \geq 0$ and $\alpha_1^2+\alpha_2^2=1$. Let $K=\|H_1\|$. Then we have,
\begin{eqnarray*}
\bra{v} H \ket{v} &\geq& \bra{v} H_1 \ket{v} + J \alpha_2^2 \\
         &=& (1-\alpha_2^2) \bra{v_1}H_1 \ket{v_1} + 2 \alpha_1 \alpha_2 {\rm Re} \bra{v_1} H_1
               \ket{v_2}+\alpha_2^2 \bra{v_2} H_1 \ket{v_2} + J \alpha_2^2 \\
         &\geq& \bra{v_1}H_1 \ket{v_1} - K \alpha_2^2 - 2 K \alpha_2
                - K \alpha_2^2 + J \alpha_2^2 \\
         &=& \bra{v_1}H_1 \ket{v_1} + (J-2K) \alpha_2^2 - 2 K \alpha_2 \\
         &\ge& \lambda(H_1|_\calS) + (J-2K) \alpha_2^2 - 2 K \alpha_2
\end{eqnarray*}
where we used $\alpha_1^2 = 1-\alpha_2^2$ and $\alpha_1 \le 1$. Since $(J-2K) \alpha_2^2 - 2 K \alpha_2$ is minimized
for $\alpha_2=K/(J-2K)$, we have
$$ \bra{v} H \ket{v} \geq \lambda(H_1|_\calS) - \frac{K^2}{J-2K}.$$
\end{proof}

\section{Kitaev's Construction}\label{sec:previous}

In this section we reprove Kitaev's result that $O(\log n)$-{\locHam} is $\QMA$-complete.
The difference between our version of the proof and
the original one in \cite{Kitaev:book} is that we do not use their geometrical lemma to obtain the result, but rather
apply our Lemma \ref{lem:leak}. This paves the way to the later proof that $2$-{\locHam} is $\QMA$-complete.

\begin{figure}[h]
\center{
 \epsfxsize=4in
 \epsfbox{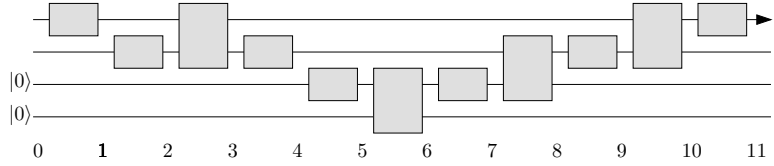}}
 \caption{A circuit with $T=11$, $N=4$ and $m=2$.}
 \label{fig:circuit1}
\end{figure}

As mentioned before, the proof that $O(\log n)$-{\locHam} is in $\QMA$ appears in \cite{Kitaev:book}.
Hence, our goal is to show that any problem in $\QMA$ can be reduced to $O(\log n)$-{\locHam}. Let $V_x=V(\ket{x},\cdot )=U_T\cdots U_1$
be a quantum verifier circuit of size $T=\poly(|x|)$ operating on $N=\poly(|x|)$ qubits.\footnote{For ease of notation we
hardwire the dependence on the input $x$ into the circuit.} Here and in what follows later we assume without loss of
generality that each $U_i$ is either a one-qubit gate or a two-qubit gate.
We further assume that $T\ge N$ and that initially, the first $m=p(|x|)$ qubits contain the proof
and the remaining ancillary $N-m$ qubits are zero (see Definition \ref{Def:QMA}). Finally, we assume that the output of
the circuit is written into the first qubit (i.e., it is $|1 \ra$ if the circuit accepts). See Figure
\ref{fig:circuit1}.

The constructed Hamiltonian $H$ operates on a space of $n=N+\log (T+1)$ qubits. The first $N$ qubits represent the computation
and the last $\log (T+1)$ qubits represent the possible values $0,\ldots,T$ for the clock:
\begin{equation*}
H =  H_{out} +J_{in}H_{in} +  J_{prop}H_{prop}.
\end{equation*}
The coefficients $J_{in}$ and $J_{prop}$ will be chosen later to be some large polynomials in $N$. The terms are given
by
 \begin{eqnarray}\label{eq:Kitaevterms}
 &&H_{in} = \sum_{i=m+1}^{N} \ketbra{1}_i \otimes \ketbra{0} \quad \quad \quad
 H_{out} = (T+1) \ketbra{0}_1 \otimes \ketbra{T}  \nonumber\\
 &&H_{prop} = \sum_{t=1}^{T} H_{prop,t}
 \end{eqnarray}
and
\begin{equation} \label{eq:Hprop}
H_{prop,t} =  \frac{1}{2} \left( I \otimes \ketbra{t}
                + I \otimes \ketbra{t\mns 1}
                - U_t \otimes \ket{t} \bra{t\mns 1}
                - U_t^\dag \otimes \ket{t\mns 1} \bra{t}
                 \right)
\end{equation}
for $1 \leq t \leq T$ where $\ketbra{\alpha}_i$ denotes the projection on the subspace in which the $i$'th qubit is
$\ket{\alpha}$. It is understood that the first part of each tensor product acts on the space of the $N$ computation
qubits and the second part acts on the clock qubits. $U_t$ and $U_t^\dagger$ in $H_{prop,t}$ act on the same
computational qubits as $U_t$ does when it is employed in the verifier's circuit $V_x$. Intuitively, each Hamiltonian
`checks' a certain property by increasing the eigenvalue if the property doesn't hold: The Hamiltonian $H_{in}$ checks
that the input of the circuit is correct (i.e., none of the last $N-m$ computation qubits is $1$), $H_{out}$ checks
that the output bit indicates acceptance and $H_{prop}$ checks that the propagation is according to the circuit. Notice
that these Hamiltonians are $O(\log n)$-local since there are $\log (T+1)=O(\log n)$ clock qubits.

To show that a problem in $\QMA$ reduces to the $O(\log n)$-{\locHam} problem with $H$ chosen as above, we prove the
following lemma.

\begin{lemma}
If the circuit $V_x$ accepts with probability more than $1-\eps$ on some input $\ket{\xi,0}$, then the Hamiltonian $H$
has an eigenvalue smaller than $\eps$. If the circuit $V_x$ accepts with probability less than $\eps$ on all inputs
$\ket{\xi,0}$, then all eigenvalues of $H$ are larger than $\frac{3}{4}-\eps$.
\end{lemma}


\begin{proof}
Assume the circuit $V_x$ accepts with probability more than $1-\eps$ on some $\ket{\xi,0}$. Define
\begin{equation*}
 \ket{\eta} = \frac{1}{\sqrt{T+1}}
   \sum_{t=0}^T U_t\cdots U_1 \ket{\xi,0} \otimes \ket{t}.
\end{equation*}
It can be seen that $\bra{\eta} H_{prop} \ket{\eta} = \bra{\eta} H_{in} \ket{\eta} = 0$ and that $\bra{\eta} H_{out}
\ket{\eta} < \eps$. Hence, the smallest eigenvalue of $H$ is less than $\eps$. It remains to prove the second part of
the lemma. So now assume the circuit $V_x$ accepts with probability less than $\eps$ on all inputs $|\xi,0\ra$.

Let $\calS_{prop}$ be the groundspace of the Hamiltonian $H_{prop}$. It is easy to see that $\calS_{prop}$ is a
$2^N$-dimensional space whose basis is given by the states
\begin{equation}\label{eq:eta}
 \ket{\eta_i} = \frac{1}{\sqrt{T+1}}
   \sum_{t=0}^T U_t\cdots U_1 \ket{i} \otimes \ket{t}
\end{equation}
where $i \in \{0,\ldots ,2^N-1\}$ and $\ket{i}$ represents the $i$th vector in the computational basis on the $N$
computation qubits. These states have eigenvalue $0$. The states in $\calS_{prop}$ represent the correct propagation
from an initial state on the $N$ computation qubits according to the verifier's circuit $V_x$.

We would like to apply Lemma \ref{lem:leak} with the space ${\cal S}_{prop}$. For that, we need to establish that
$J_{prop} H_{prop}$ gives a sufficiently large ($\poly(N)$) penalty to states in ${\cal S}^\perp_{prop}$. In other
words, the smallest non-zero eigenvalue of $H_{prop}$ has to be lower bounded by some inverse polynomial in $N$. This
has been shown in \cite{Kitaev:book}, but we wish to briefly recall it here, as it will apply in several instances
throughout this paper.

\begin{claim}[\cite{Kitaev:book}]\label{claim:gap}
The smallest non-zero eigenvalue of $H_{prop}$ is at least $c/T^2$
for some constant $c>0$.
\end{claim}
\begin{proof}
We first apply the change of basis
\begin{equation*}
W=\sum_{t=0}^T U_t \cdots U_1 \otimes \ket{t}\bra{t}
\end{equation*}
which transforms $H_{prop}$ to
\begin{equation*}
W^\dagger H_{prop} W =\sum_{t =1}^T
 I \otimes \frac{1}{2}\left(\ket{t}\bra{ {t}} +  \ket{ {t\mns 1}}\bra{ {t\mns 1}}- \ket{ t}\bra{t\mns 1}- \ket{t\mns 1}\bra{t} \right).
\end{equation*}
The eigenspectrum of $H_{prop}$ is unchanged by this transformation. The resulting Hamiltonian is block-diagonal with $2^N$ blocks of size $T+1$.
\begin{eqnarray}\label{eq:h}
W^\dagger H_{prop} W & = & I \otimes \left(
\begin{array}{rrrrrrr}
\smfrac{1}{2} & -\smfrac{1}{2} &0 & & \cdots& & 0 \\ -\smfrac{1}{2} & 1 & -\smfrac{1}{2} & 0 & \ddots & & \vdots\\ 0 &
-\smfrac{1}{2} & 1 & -\smfrac{1}{2} & 0 & \ddots & \vdots\\ & \ddots & \ddots & \ddots & \ddots & \ddots & \\ \vdots& &
0 & -\smfrac{1}{2} &1 & -\smfrac{1}{2}& 0 \\ & & & 0 & -\smfrac{1}{2} &1 & -\smfrac{1}{2} \\ 0& & \cdots& & 0&
-\smfrac{1}{2} & \smfrac{1}{2} \\
\end{array}
\right).
\end{eqnarray}
Using standard techniques, one can show that the smallest non-zero eigenvalue of each $(T+1) \times (T+1)$ block matrix
is bounded from below by $c/T^2$, for some constant $c>0$.
\end{proof}

Hence any eigenvector of $J_{prop}H_{prop}$ orthogonal to $\calS_{prop}$ has eigenvalue at least $J = c J_{prop}/T^2$.
Let us apply Lemma \ref{lem:leak} with
 \begin{align*}
 H_1&=H_{out}+J_{in}H_{in}  &    H_2 &= J_{prop}H_{prop}.
 \end{align*}
Note that $\|H_1\| \leq \|H_{out}\|+J_{in} \|H_{in}\|\le T+1+J_{in} N \le \poly(N)$ since $H_{in}$ and $H_{out}$ are
sums of orthogonal projectors and $J_{in}=\poly(N)$. Lemma \ref{lem:leak} implies that we can choose
$J_{prop}=JT^2/c=\poly(N)$, such that  $\lambda(H)$ is lower bounded by $\lambda(H_1|_{\calS_{prop}})-\frac{1}{8}$.
With this in mind, let us now consider the Hamiltonian $H_{1}|_{\calS_{prop}}$ on $\calS_{prop}$.

Let $\calS_{in} \subset \calS_{prop}$ be the groundspace of $H_{in}|_{\calS_{prop}}$. Then $\calS_{in}$ is a
$2^m$-dimensional space whose basis is given by states as in Eq. (\ref{eq:eta}) with $\ket{i}=\ket{j,0}$, where $\ket{j}$ is
a computational basis state on the first $m$ computation qubits. We apply Lemma \ref{lem:leak} again {\em inside}
$\calS_{prop}$ with
 \begin{align*}
 H_1 &= H_{out}|_{\calS_{prop}}  &   H_2&=J_{in} H_{in}|_{\calS_{prop}}.
 \end{align*}
This time, $\|H_1\| \le \|H_{out}\| =T+1=\poly(N)$. Any eigenvector of $H_2$ orthogonal to $\calS_{in}$ inside $\calS_{prop}$ has
eigenvalue at least $J_{in}/(T+1)$. Hence, there is a $J_{in}=\poly(N)$ such that $\lambda(H_1+H_2)$ is lower bounded by
$\lambda(H_{out}|_{\calS_{in}})-\frac{1}{8}$.

Since the circuit $V_x$ accepts with probability less than $\eps$ on all inputs $|\xi,0\ra$, we have that all
eigenvalues of $H_{out}|_{\calS_{in}}$ are larger than $1-\eps$. Hence the smallest eigenvalue of $H$ is larger than
$1-\eps-\frac{2}{8} = \frac{3}{4}-\eps$, proving the second part of the lemma.
\end{proof}

\section{The 2-local Construction}\label{sec:2local}

\paragraph{Previous constructions:}
Let us give an informal description of ideas used in previous improvements on Kitaev's construction;
these ideas will also appear in our proof.
The first idea is to represent the clock register
in {\em unary notation}. Then, the clock register consists of $T$ qubits and time step $t \in \{0,\ldots,T\}$ is represented
by $\ket{1^t 0^{T-t}}$. The crucial observation is that clock terms that used to involve
$\log (T+1)$ qubits, can now be replaced by $3$-local terms that are essentially
equivalent. For example, a term like $\ket{t\mns 1} \bra{t}$ can be replaced
by the term $\ket{100} \bra{110}_{t-1,t,t+1}$.
Since the gates $U_t$ involve at most two qubits, we obtain a $5$-local Hamiltonian. This is essentially the
way $5$-{\locHam} was shown to be $\QMA$-complete in \cite{Kitaev:book}.
The only minor complication is that we need to get rid of illegal
clock states (i.e., ones that are not a unary representation).
This is done by the addition of a ($2$-local) Hamiltonian $H_{clock}$ that penalizes
a clock state whenever $1$ appears after $0$.

This result was further improved to $3$-{\locHam} in \cite{Kempe:03a}.
The main idea there is to replace a $3$-local clock term like $\ket{100} \bra{110}_{t-1,t,t+1}$
by the $1$-local term $\ket{0}\bra{1}_t$. These one-qubit terms are no longer
equivalent to the original clock terms. Indeed, it can be seen that they have unwanted transitions into
illegal clock states. The main idea in \cite{Kempe:03a} was that by
giving a large penalty to illegal clock states (i.e., by multiplying $H_{clock}$ by
some large number) and applying the projection lemma, we can
essentially project these one-qubit terms to the subspace of legal clock states.
Inside this subspace, these terms become the required clock terms.

\begin{figure}[h]
\center{
 \epsfxsize=2in
 \epsfbox{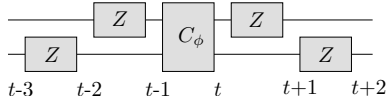}}
 \caption{A modified $\cphase$ gate applied at step $t$}
 \label{fig:cphase}
\end{figure}

\paragraph{The 2-local construction:}
Most of the terms that appear in the construction of \cite{Kempe:03a} are already
$2$-local. The only $3$-local terms are terms as in Eq. \eqref{eq:Hprop} that correspond to two-qubit gates
(those corresponding to one-qubit gates are already $2$-local).
Hence, in order to prove our main theorem, it is enough to find a $2$-local Hamiltonian
that checks for the correct propagation
of $2$-qubit gates.
This seems difficult because the Hamiltonian must somehow couple
two computation qubits to a clock qubit. We circumvent this problem
in the following manner.
First, we isolate from the propagation Hamiltonian those terms
that correspond to one-qubit gates and we multiply
these terms by some large factor.
Using the projection lemma, we can project the remaining Hamiltonians into a
space where the $1$-qubit-gate propagation is correct. In other words,
at this stage we can assume that our space is spanned by states
that correspond to legal propagation according to the $1$-qubit gates.
This allows us to couple clock qubits {\em instead} of computation qubits.
To see this, consider the circuit in Fig. \ref{fig:cphase} at time
$t$ and at time $t+2$. A $Z$ gate flips the phase of a qubit if its state is $\ket{1}$
and leaves it unchanged otherwise. Hence, the phase difference between time $t$
and time $t+2$ corresponds to the parity of the two qubits.
This phase difference can be detected by a $2$-local term such
as $\ket{00}\bra{11}_{t+1, t+2}$.
The crucial point here is that by using a term involving only two clock qubits,
we are able to check the state of two computation qubits (in this case, their parity)
at a certain time. This is the main idea in our proof.

\paragraph{}
We now present the proof of the main theorem in detail. We start by making some further assumptions on the circuit $V_x$, all
without loss of generality.
First, we assume that in addition to one-qubit gates, the circuit contains only the controlled phase gate, $\cphase$.
This two-qubit gate is diagonal in the computational basis and flips the sign of the state $\ket{11}$,
\begin{align*}
\cphase = \cphase^\dagger = \ket{00}\bra{00}+\ket{01}\bra{01}+\ket{10}\bra{10}-\ket{11}\bra{11}.
\end{align*}
It is known \cite{Barenco:95a,Nielsen:book} that quantum circuits consisting of one-qubit gates and $\cphase$ gates are
universal\footnote{The original universal gate set in \cite{Barenco:95a} consists of one-qubit gates and {\sc
CNOT} gates. It is, however, easy to see that a {\sc CNOT} gate can be obtained from a $\cphase$ gate by conjugating the
second qubit with Hadamard gates (see \cite{Nielsen:book}).} and can simulate any other quantum circuit with only
polynomial overhead. Second, we assume that each $\cphase$ gate is both preceded and followed by two $Z$ gates,
one on each qubit, as in Figure \ref{fig:cphase}.
The $Z$ gate is defined by $\ketbra{0} - \ketbra{1}$; i.e., it is a diagonal one-qubit
gate that flips the sign of $|1\ra$. Since both the $Z$ gate and the $\cphase$ gate are diagonal, they commute and the
effect of the $Z$-gates cancels out. This assumption makes the circuit at most five times
bigger. Finally, we assume that the $\cphase$ gates are applied at
regular intervals. In other words, if $T_2$ is the number of $\cphase$ gates and $L$ is the interval length, then a
$\cphase$ gate is applied at steps $L,2L,\ldots, T_2L$. Before the first $\cphase$ gate, after the last $\cphase$ gate
and between any two consecutive $\cphase$ gates we have $L-1$ one-qubit gates. This makes the total number of gates in
the resulting circuit $T=(T_2+1)L-1$.

 We construct a Hamiltonian $H$ that operates on a space of $N+T$ qubits. The first $N$ qubits represent the computation and the last $T$ qubits represent the clock. We think of the clock as represented in unary,
\begin{equation}\label{eq:unary}
\ket{\widehat{t}} \defeq \ket{\underbrace{1\ldots1}_t\underbrace{0\ldots0}_{T-t}}.
\end{equation}
Let $T_1$ be the time steps in which a one-qubit gate is applied. Namely, $T_1 = \{1,\ldots, T\} \setminus
\{L,2L,\ldots, T_2L\}$. Then
\begin{equation*}
H = H_{out} + J_{in} H_{in}  + J_{2}H_{prop2} + J_{1}H_{prop1} + J_{clock}H_{clock},
\end{equation*}
where
 \begin{align*}
&H_{in} = \sum_{i=m+1}^{N} \ketbra{1}_i \otimes \ketbra{0}_1 \quad \quad \quad H_{out} = (T+1)\ketbra{0}_1 \otimes \ketbra{1}_T  \\
 &H_{clock} = \sum_{1\le i < j \le T} I \otimes \ketbra{01}_{ij}.
\end{align*}
The terms $H_{prop1}$ and $H_{prop2}$, which represent the correct propagation according to the
$1$-qubit gates and $2$-qubit gates respectively, are defined as:
\begin{align*}
 H_{prop1} &= \sum_{t \in T_1} H_{prop,t}   &   H_{prop2} &= \sum_{l=1}^{T_2} \left( H_{qubit,lL}+H_{time,lL} \right) \nonumber
\end{align*}
with
\begin{eqnarray*}
 H_{prop,t} &=& \frac{1}{2} \left(I \otimes \ketbra{10}_{t,t+1}
                + I \otimes \ketbra{10}_{t-1,t}
                - U_t \otimes \ket{1}\bra{0}_t
                - U_t^\dag \otimes \ket{0}\bra{1}_t
                \right) \nonumber
 \end{eqnarray*}
for $t \in T_1 \cap \{2, \ldots , T-1\}$ and
 \begin{eqnarray*}
 H_{prop,1} &=& \frac{1}{2} \left(I \otimes \ketbra{10}_{1,2}
                + I \otimes \ketbra{0}_{1}
                - U_1 \otimes \ket{1}\bra{0}_1
                - U_1^\dag \otimes \ket{0}\bra{1}_1
                \right) \\
 H_{prop,T} &=& \frac{1}{2} \left(I \otimes \ketbra{1}_{T}
                + I \otimes \ketbra{10}_{T-1,T}
                - U_T \otimes \ket{1} \bra{0}_T
                - U_T^\dag \otimes \ket{0}\bra{1}_T
                \right)
 \end{eqnarray*}
and, with $f_t$ and $s_t$ being the first and second qubit of the $\cphase$ gate at time $t$,
\begin{align*}
H_{qubit,t} &= \frac{1}{2}\left( -2 \ketbra{0}_{f_t} - 2 \ketbra{0}_{s_t} + \ketbra{1}_{f_t} + \ketbra{1}_{s_t} \right)
   \otimes \left( \ket{1}\bra{0}_{t} + \ket{0} \bra{1}_{t}\right) \\
H_{time,t} &= \frac{1}{8} I \otimes \left( \ketbra{10}_{t,t+1} +6 \ketbra{10}_{t+1,t+2} +\ketbra{10}_{t+2,t+3} \right. \\
  &\qquad\qquad +  2\ket{11}\bra{00}_{t+1,t+2}+ 2\ket{00}\bra{11}_{t+1,t+2}  \\
  &\qquad\qquad +   \ket{1}\bra{0}_{t+1}+  \ket{0} \bra{1}_{t+1}+ \ket{1}\bra{0}_{t+2}+  \ket{0} \bra{1}_{t+2}  \\
  &\qquad\qquad +    \ketbra{10}_{t-3,t-2} +6\ketbra{10}_{t-2,t-1} +\ketbra{10}_{t-1,t} \\
  &\qquad\qquad +   2\ket{11}\bra{00}_{t-2,t-1}+ 2\ket{00}\bra{11}_{t-2,t-1}  \\
  &\qquad\qquad +    \left. \ket{1}\bra{0}_{t-2}+  \ket{0} \bra{1}_{t-2}+ \ket{1} \bra{0}_{t-1}+  \ket{0}\bra{1}_{t-1} \right) .
\end{align*}
At this point, these last two expressions might look strange. Let us say that later, when we consider their restriction
to a smaller space, the reason for this definition should become clear. Note that all the above terms are at most
$2$-local. We will later choose $J_{in} \ll J_2 \ll J_1 \ll J_{clock} \le \poly(N)$. As in Section \ref{sec:previous}, we
have to prove the following lemma:
\begin{lemma}
Assume that the circuit $V_x$ accepts with probability more than $1-\eps$ on some input $\ket{\xi,0}$. Then $H$ has an
eigenvalue smaller than $\eps$.  If the circuit $V_x$ accepts with probability less than $\eps$ on all inputs
$\ket{\xi,0}$, then all eigenvalues of $H$ are larger than $\frac{1}{2}-\eps$.
\end{lemma}
\begin{proof}
If the circuit $V_x$ accepts with probability more than $1-\eps$ on some input $\ket{\xi,0}$ then the state
\begin{equation*}
 \ket{\eta} = \frac{1}{\sqrt{T+1}}
   \sum_{t=0}^T U_t\cdots U_1 \ket{\xi,0} \otimes \ket{\widehat{t}}
\end{equation*}
satisfies $\bra{\eta} H \ket{\eta} \le \eps$. In order to see this, one can check that
$$ \bra{\eta} H_{clock} \ket{\eta} = \bra{\eta} H_{prop1} \ket{\eta} = \bra{\eta} H_{prop2} \ket{\eta}
 = \bra{\eta} H_{in}\ket{\eta} = 0$$
and $\bra{\eta} H_{out} \ket{\eta} \le \eps$. However, verifying that $\bra{\eta} H_{prop2} \ket{\eta}=0$ can be quite
tedious. Later in the proof, we will mention an easier way to see this.

In the following, we will show that if the circuit $V_x$ accepts with probability less than $\eps$ on all inputs
$\ket{\xi,0}$, then all eigenvalues of $H$ are larger than $\frac{1}{2}-\eps$. The proof of this is based on four
applications of Lemma \ref{lem:leak}. Schematically, we proceed as follows:
\begin{equation*}
{\cal H} \supset \calS_{legal} \supset \calS_{prop1} \supset \calS_{prop} \supset \calS_{in}
\end{equation*}
where $\calS_{legal}$ corresponds to states with {\em legal} clock states written in unary, and $\calS_{prop1}$ is
spanned by states in the legal clock space whose propagation at time steps corresponding to {\em one-qubit} gates (that is, in
$T_1$) is correct. Finally, $\calS_{prop}$ and $\calS_{in}$ are defined in almost the same way as in Section
\ref{sec:previous}. These spaces will be described in more detail later.

\paragraph {Norms:}
Note that all relevant norms, as needed in Lemma \ref{lem:leak}, are polynomial in $N$. Indeed, we have
$\|H_{out}\|=T+1$ and $\|H_{in}\| \leq N$ as in Section \ref{sec:previous}, $\|H_{prop1}\| \leq \sum_{t \in T_1}
\|H_{prop,t}\| \leq 2T$ (each term in $H_{prop1}$ has norm at most $2$) and $\|H_{prop2}\|\leq \sum_{t=1}^{ T_2}
(\|H_{qubit,lL}\|+\|H_{time,lL}\|) \leq O(T_2) \leq O(T)$.

\paragraph{1. Restriction to legal clock states in $\calS_{legal}$:}
Let $\calS_{legal}$ be the $(T+1)2^N$-dimensional space spanned by states with a legal unary representation on the $T$
clock qubits, i.e., by states of the form $|\widetilde{\xi}\ra \otimes |\widehat {t}\ra$ with $ |\widehat {t}\ra$ as in
Eq. (\ref{eq:unary}). In this first stage we apply Lemma \ref{lem:leak} with
 \begin{align*}
 H_1&=H_{out}+J_{in}H_{in}+ J_{2}H_{prop2} + J_{1}H_{prop1}  & H_2&=J_{clock}H_{clock}.
 \end{align*}
Notice that $\calS_{legal}$ is an eigenspace of $H_2$ of eigenvalue $0$ and that states orthogonal to $\calS_{legal}$
have eigenvalue at least $J_{clock}$. Lemma \ref{lem:leak} implies that we can choose
$J_{clock}=\poly(\|H_1\|)=\poly(N)$ such that $\lambda(H)$ can be lower bounded by
$\lambda(H_1|_{\calS_{legal}})-\frac{1}{8}$. Hence, in the remainder of the proof, it is enough to study
$H_{1}|_{\calS_{legal}}$ inside the space $\calS_{legal}$. This can be written as:
\begin{equation*}
H_{out}|_{\calS_{legal}} + J_{in} H_{in}|_{\calS_{legal}}  + J_{2}H_{prop2}|_{\calS_{legal}} + J_{1}H_{prop1}|_{\calS_{legal}}
\end{equation*}
with
\begin{align*} 
 H_{in}|_{\calS_{legal}} &= \sum_{i=m+1}^{N} \ketbra{1}_i \otimes \ket{\widehat 0}\bra{\widehat 0} \quad \quad \quad
  H_{out}|_{\calS_{legal}} = (T+1)\ketbra{0}_1 \otimes \ketbra{\widehat T}  \nonumber \\
 H_{prop,t}|_{\calS_{legal}} &= \frac{1}{2} \left(I \otimes \ketbra{\widehat t}
                + I \otimes \ketbra{\widehat{t\mns 1}}
                - U_t \otimes \ket{\widehat t} \bra{\widehat{t\mns 1}}
                - U_t^\dag \otimes \ket{\widehat{t\mns 1}} \bra{\widehat t}
                \right) \nonumber \\
 H_{qubit,t}|_{\calS_{legal}} &=  \frac{1}{2}  \left( - 2\ketbra{0}_{f_t}-2\ketbra{0}_{s_t}+\ketbra{1}_{f_t}+\ketbra{1}_{s_t}      \right)\otimes  \left(\ket{\widehat t} \bra{\widehat{t\mns 1}}+ \ket{\widehat{t\mns 1}} \bra{\widehat t}\right)\nonumber \\
 H_{time,t}|_{\calS_{legal}} &= \frac{1}{8}  I \otimes \left( \ketbra{\widehat {t}} + 6\ketbra{\widehat {t\pls 1}} +\ketbra{\widehat {t\pls 2}}  \right. \nonumber \\
  &\qquad \qquad +    2\ket{\widehat {t\pls 2}} \bra{\widehat {t}} +2\ket{\widehat {t}}\bra{\widehat {t\pls 2}} +\ket{\widehat {t\pls 1}} \bra{\widehat {t}} + \ket{\widehat {t}}\bra{\widehat {t\pls 1}}+\ket{\widehat {t\pls 2}} \bra{\widehat {t\pls 1}} + \ket{\widehat {t\pls 1}}\bra{\widehat {t\pls 2}} \nonumber \\
  &\qquad \qquad +    \ketbra{\widehat {t\mns 3}} + 6 \ketbra{\widehat {t\mns 2 }} +\ketbra{\widehat {t\mns 1}}  \nonumber \\
  &\qquad \qquad \left. +   2\ket{\widehat {t\mns 1}} \bra{\widehat {t\mns 3}} +2\ket{\widehat {t\mns
     3}}\bra{\widehat {t\mns 1}} +\ket{\widehat {t\mns 2 }} \bra{\widehat {t\mns 3}} + \ket{\widehat {t\mns 3}}\bra{\widehat
     {t\mns 2 }}+\ket{\widehat {t\mns 1}} \bra{\widehat {t\mns 2 }} + \ket{\widehat {t\mns 2 }}\bra{\widehat {t\mns 1}}
     \right).
\end{align*}
The above was obtained by noting that the projection of a term like, say, $\ketbra{10}_{t,t+1}$ on $\calS_{legal}$ is
exactly $\ketbra{\hat{t}}$. Similarly, the projection of the term $\ket{1}\bra{0}_{t + 1}$ is
$\ket{\widehat{t \pls 1}}\bra{\hat{t}}$.\footnote{Notice that we do not have terms like $\ketbra{1}_{t}$; its projection on
$\calS_{legal}$ is not $\ketbra{\hat t}$ but rather $\ketbra{\hat t}+\cdots+\ketbra{\widehat T}$.} By rearranging
terms, the above expression can be written as a sum of projectors:
\begin{align}\label{eq:Htime}
H_{time,t}|_{\calS_{legal}} &= \frac{1}{8} I \otimes \left\{ 2  \left(\ket{\widehat t}+\ket{\widehat {t\pls 1}}\right)\left(\bra{\widehat t}+\bra{\widehat {t\pls 1}}\right)+2\left(\ket{\widehat {t\pls 1}}+\ket{\widehat {t\pls 2}}\right)\left(\bra{\widehat {t\pls 1}}+\bra{\widehat {t\pls 2}}\right) \right. \nonumber \\
   &\qquad \qquad +  \left(\ket{\widehat t}-\ket{\widehat {t\pls 1}}\right) \left(\bra{\widehat t}-\bra{\widehat {t\pls 1}}\right) +\left(\ket{\widehat {t\pls 1}}-\ket{\widehat {t\pls 2}}\right) \left(\bra{\widehat {t\pls 1}}-\bra{\widehat {t\pls 2}}\right) \nonumber \\
   &\qquad \qquad -  2\left(\ket{\widehat {t}}-\ket{\widehat {t\pls 2}}\right)\left(\bra{\widehat {t}}-\bra{\widehat {t\pls 2}}\right)\nonumber \\
   &\qquad \qquad +   2 \left(\ket{\widehat {t\mns 3}}+\ket{\widehat {t\mns 2 }}\right)\left(\bra{\widehat {t\mns 3}}+\bra{\widehat {t\mns 2 }}\right)+2\left(\ket{\widehat {t\mns 2 }}+\ket{\widehat {t\mns 1}}\right)\left(\bra{\widehat {t\mns 2 }}+\bra{\widehat {t\mns 1}}\right)  \nonumber \\
   &\qquad \qquad +   \left(\ket{\widehat {t\mns 3}}-\ket{\widehat {t\mns 2 }}\right) \left(\bra{\widehat {t\mns 3}}-\bra{\widehat {t\mns 2 }}\right) +\left(\ket{\widehat {t\mns 2 }}-\ket{\widehat {t\mns 1}}\right) \left(\bra{\widehat {t\mns 2 }}-\bra{\widehat {t\mns 1}}\right) \nonumber \\
   &\qquad \qquad -  2\left. \left(\ket{\widehat {t\mns 3}}-\ket{\widehat {t\mns 1}}\right)\left(\bra{\widehat {t\mns 3}}-\bra{\widehat {t\mns 1}}\right) \right\}.
\end{align}
Notice that the above expression is symmetric around $t-\frac{1}{2}$ (i.e., switching $t-1$ with $t$, $t-2$ with $t+1$,
and $t-3$ with $t+2$ does not change the expression). Let us also mention that the fact that we have terms like
$\ket{\widehat {t}}-\ket{\widehat {t\pls 2}}$ is crucial in our proof. They allow us to compare the state at time $t$
to the state at time $t+2$.

\paragraph{2. Restriction to $\calS_{prop1}$:}
We now apply Lemma \ref{lem:leak} inside $\calS_{legal}$ with
 \begin{align*}
 H_1&=\left( H_{out}+ J_{in}H_{in}+J_2 H_{prop2}\right)|_{\calS_{legal}}  & H_2&=J_1H_{prop1}|_{\calS_{legal}}.
 \end{align*}
Let $\calS_{prop1}$ be the $2^N (T_2+1)$-dimensional space given by all states that represent correct propagation on
all one-qubit gates. More precisely, let
\begin{equation}\label{eq:etasingle}
\ket{\eta_{l,i}} \defeq \frac{1}{\sqrt{L}}\sum_{t = l L}^{(l+1)L-1} U_{t} \cdots U_1 \ket{i} \otimes |\widehat t\ra,
\end{equation}
where $l\in \{0,\ldots,T_2\}$, $i \in \{0,\ldots,2^N-1\}$ and $\ket{i}$ represents the $i$th vector in the
computational basis. Then these states form a basis of $\calS_{prop1}$. It is easy to see that each $\ket{\eta_{l,i}}$
is an eigenvector of $H_{prop1}$ of eigenvalue $0$. Hence, $\calS_{prop1}$ is an eigenspace of eigenvalue $0$ of
$H_{prop1}|_{\calS_{legal}}$. Furthermore, $H_{prop1}|_{\calS_{legal}}$ decomposes into $T_2+1$ invariant blocks, with
the $l$th block spanned by states of the form $U_{t} \cdots U_{1} \ket{i} \otimes |\widehat t\ra$ for $t=lL, \ldots ,
(l+1)L-1$. Inside such a block $H_{prop1}|_{\calS_{legal}}$ corresponds exactly to $H_{prop}$ of Section
\ref{sec:previous}, Eqs. (\ref{eq:Kitaevterms},\ref{eq:Hprop}). By Claim \ref{claim:gap}, its non-zero eigenvalues are at
least $c/L^2 \geq c/T^2$ for some constant $c>0$ and hence the smallest non-zero eigenvalue of
$H_{prop1}|_{\calS_{legal}}$ is also at least $c/T^2$. Therefore, all eigenvectors of $H_2$ orthogonal to
$\calS_{prop1}$ have eigenvalue at least $J=J_1c/T^2$ and Lemma \ref{lem:leak} implies that for $J_{1} \ge \poly(N)$,
$\lambda(H_1+H_2)$ can be lower bounded by $\lambda(H_{1}|_{\calS_{prop1}}) - \frac{1}{8}$.

Hence, in the remainder of the proof, it is enough to study
$$H_{out}|_{\calS_{prop1}}+ J_{in}H_{in}|_{\calS_{prop1}}+J_2 H_{prop2}|_{\calS_{prop1}}.$$
Let us find $H_{prop2}|_{\calS_{prop1}}$. Let $t=lL$ be the time at which the $l$th $\cphase$ gate is applied and
consider the projection of a state $\ket{\eta_{l,i}}$ onto the space spanned by the computation qubits and
$|\widehat{t}\ra, |\widehat{t\pls 1}\ra, |\widehat{t\pls 2}\ra$. Since at time $t+1$ (resp., $t+2$) a $Z$ gate is
applied to qubit $f_t$ (resp., $s_t$), this projection is a linear combination of the following four states:
\begin{eqnarray*} \label{eq:legalsprop1}
 &&|00\ra _{f_t,s_t} |\xi_{00}\ra \otimes \left(\ket{\widehat t}+\ket{\widehat{t\pls 1}}+\ket{\widehat{t\pls 2}}\right)\nonumber \\
 &&|01\ra _{f_t,s_t} |\xi_{01}\ra \otimes \left(\ket{\widehat t}+\ket{\widehat{t\pls 1}}-\ket{\widehat{t\pls 2}}\right)\nonumber \\
 &&|10\ra _{f_t,s_t} |\xi_{10}\ra \otimes \left(\ket{\widehat t}-\ket{\widehat{t\pls 1}}-\ket{\widehat{t\pls 2}}\right)\nonumber \\
 &&|11\ra _{f_t,s_t} |\xi_{11}\ra \otimes \left(\ket{\widehat t}-\ket{\widehat{t\pls 1}}+\ket{\widehat{t\pls 2}}\right),
\end{eqnarray*}
where $\ket{\xi_{b_1b_2}}$ is an arbitrary state on the remaining $N-2$ computation qubits. This implies that the
restriction to $\calS_{prop1}$ of the projector on, say, $\ket{\widehat t}+\ket{\widehat{t\pls 1}}$ from Eq.
(\ref{eq:Htime}) is essentially the same as the restriction to $\calS_{prop1}$ of the projector on
$\ket{0}_{f_t}\ket{\widehat{t}}$. More precisely, for all $l_1,l_2,i_1,i_2$ we have
$$ \frac{1}{4} \bra{\eta_{l_1,i_1}} \Big( I \otimes \big( \ket{\widehat t}+\ket{\widehat{t\pls 1}}\big) \big(\bra{\widehat t}+\bra{\widehat{t\pls 1}}\big) \Big) \ket{\eta_{l_2,i_2}}
      = \bra{\eta_{l_1,i_1}} \left(\ket{0}\bra{0}_{f_t}\otimes \ket{\widehat{t}} \bra{\widehat{t}}\right)\ket{\eta_{l_2,i_2}}.
$$
Similarly, the term involving $\ket{\widehat t}-\ket{\widehat{t\pls 2}}$ satisfies
$$ \frac{1}{4} \bra{\eta_{l_1,i_1}} \Big( I \otimes \big( \ket{\widehat t} - \ket{\widehat{t\pls 2}}\big) \big(\bra{\widehat t} - \bra{\widehat{t\pls 2}}\big) \Big) \ket{\eta_{l_2,i_2}}
      = \bra{\eta_{l_1,i_1}} \left( \big( \ketbra{01}_{f_t,s_t} + \ketbra{10}_{f_t,s_t} \big) \otimes \ket{\widehat{t}} \bra{\widehat{t}}\right)\ket{\eta_{l_2,i_2}}.
$$
Observe that the right-hand side involves two computation qubits and the clock register.
Being able to obtain such a term from two-local terms is a crucial ingredient in this proof.

Following a similar calculation, we see that from the terms involving $|\widehat{t\mns1}\ra, |\widehat{t\mns 2}\ra, |\widehat{t\mns 3}\ra$ we obtain
projectors involving $\ket{\widehat{t \mns 1}}$. To summarize, instead of considering $H_{time,t}|_{\calS_{prop1}}$ we
can equivalently consider the restriction to $\calS_{prop1}$ of
\begin{eqnarray*}
&& \frac{1}{2}\left( 2 \ketbra{0}_{f_t}  + 2 \ketbra{0}_{s_t}
   +  \ketbra{1}_{f_t} + \ketbra{1}_{s_t} \nonumber \right.
    -  2\left. \ketbra{01}_{f_t,s_t} - 2\ketbra{10}_{f_t,s_t} \right) \nonumber \\
    &&\qquad \qquad \otimes \left( \ketbra{\widehat{t\mns 1}} + \ketbra{\widehat{t}} \right).
\end{eqnarray*}
We now add the terms in $H_{qubit,t}$. A short calculation shows that
$\left(H_{time,t}+H_{qubit,t}\right)|_{\calS_{prop1}}$ is the same as the restriction to $\calS_{prop1}$ of
\begin{eqnarray*}
\ketbra{00}_{f_t,s_t} &\otimes& 2\left( \ket{\widehat{t\mns 1}} - \ket{\widehat{t}} \right)
        \left( \bra{\widehat{t\mns 1}} - \bra{\widehat{t}} \right)  +\\
\ketbra{01}_{f_t,s_t} &\otimes& \frac{1}{2} \left( \ket{\widehat{t\mns 1}} - \ket{\widehat{t}} \right)
        \left( \bra{\widehat{t\mns 1}} - \bra{\widehat{t}} \right)  +\\
\ketbra{10}_{f_t,s_t} &\otimes& \frac{1}{2}\left( \ket{\widehat{t\mns 1}} - \ket{\widehat{t}} \right)
        \left( \bra{\widehat{t\mns 1}} - \bra{\widehat{t}} \right)  +\\
\ketbra{11}_{f_t,s_t} &\otimes& ~~\left( \ket{\widehat{t\mns 1}} + \ket{\widehat{t}} \right)
        \left( \bra{\widehat{t\mns 1}} + \bra{\widehat{t}} \right) .
\end{eqnarray*}

At this point, let us mention how one can show that for the state $\ket{\eta}$ described in the beginning of this
proof, $\bra{\eta} H_{prop2} \ket{\eta}=0$. First, observe that $\ket{\eta} \in \calS_{prop1}$ (its propagation is
correct at all time steps). Next, since $\ket{\eta}$ has a $\cphase$ propagation at time $t$, the above Hamiltonian
shows that $\bra{\eta} H_{prop2} \ket{\eta}=0$.

Let us return now to the main proof. Recall that we wish to show a lower bound on the lowest eigenvalue of
 \begin{equation}\label{eq:psdbefore}
 H_{out}|_{\calS_{prop1}}+ J_{in}H_{in}|_{\calS_{prop1}}+J_2 H_{prop2}|_{\calS_{prop1}}.
 \end{equation}
In the following, we show a lower bound on the lowest eigenvalue of the Hamiltonian
 \begin{equation}\label{eq:psdafter}
  H_{out}|_{\calS_{prop1}}+ J_{in}H_{in}|_{\calS_{prop1}}+J_2 H'
 \end{equation}
 on $\calS_{prop1}$ where $H'$ satisfies that $H' \le H_{prop2}|_{\calS_{prop1}}$,
i.e., $H_{prop2}|_{\calS_{prop1}}-H'$ is positive semidefinite. Hence, any lower bound on the lowest eigenvalue of the
Hamiltonian in \eqref{eq:psdafter} implies the same lower bound on the lowest eigenvalue of the Hamiltonian in
\eqref{eq:psdbefore}. We define $H'$ as the sum over $t\in \{L,2L,\ldots, T_2L\}$ of the restriction to $\calS_{prop1}$
of
\begin{eqnarray*}\label{eq:Htime4}
\ketbra{00}_{f_t,s_t} &\otimes& \frac{1}{2}\left( \ket{\widehat{t\mns 1}} - \ket{\widehat{t}} \right)
        \left( \bra{\widehat{t\mns 1}} - \bra{\widehat{t}} \right)  +\\
\ketbra{01}_{f_t,s_t} &\otimes&  \frac{1}{2}\left( \ket{\widehat{t\mns 1}} - \ket{\widehat{t}} \right)
        \left( \bra{\widehat{t\mns 1}} - \bra{\widehat{t}} \right)  +\\
\ketbra{10}_{f_t,s_t} &\otimes&  \frac{1}{2}\left( \ket{\widehat{t\mns 1}} - \ket{\widehat{t}} \right)
        \left( \bra{\widehat{t\mns 1}} - \bra{\widehat{t}} \right)  +\\
\ketbra{11}_{f_t,s_t} &\otimes&  \frac{1}{2}\left( \ket{\widehat{t\mns 1}} + \ket{\widehat{t}} \right)
        \left( \bra{\widehat{t\mns 1}} + \bra{\widehat{t}} \right) .
\end{eqnarray*}
Equivalently, $H'$ is the sum over $t\in \{L,2L,\ldots, T_2L\}$ of
\begin{eqnarray*}\label{eq:Htime5}
\frac{1}{2} \left. \left(I \otimes \ket{\widehat{t}}\bra{\widehat{t}} + I \otimes \ket{\widehat{t \mns
1}}\bra{\widehat{t \mns 1}}-\cphase \otimes \ket{\widehat{t}}\bra{\widehat{ t \mns 1}} - C^\dagger_\phi \otimes
\ket{\widehat{t \mns 1}}\bra{\widehat{ t}}\right) \right|_{\calS_{prop1}},
\end{eqnarray*}
which resembles Eq. (\ref{eq:Hprop}). Note that this term enforces correct propagation at time step $t=lL$. We claim
that
\begin{equation}\label{eqn:hprop2}
  H'= \frac{1}{2L}\sum_{i=0}^{2^N-1} \sum_{l=1}^{T_2} \left( \ket{\eta_{l-1,i}} - \ket{\eta_{l,i}} \right)
   \left( \bra{\eta_{l-1,i}} - \bra{\eta_{l,i}} \right).
\end{equation}
The intuitive reason for this is the following. For any $i$, $\ket{\eta_{l-1,i}}+\ket{\eta_{l,i}}$ can be seen as a
correct propagation at time $t=lL$. In other words, consider the projection of $\ket{\eta_{l,i}}$ on clock
$\ket{\widehat{t}}$ and the projection of $\ket{\eta_{l-1,i}}$ on clock $\ket{\widehat{t \mns 1}}$. Then the first
state is exactly the second state after applying the $l$th $\cphase$ gate. This means that inside $\calS_{prop1}$,
checking correct propagation from time $t-1$ to time $t$ is equivalent to checking correct propagation from
$\ket{\eta_{l-1,i}}$ to $\ket{\eta_{l,i}}$.

More precisely, fix some $l$ and $t=lL$. Then, using Eq. (\ref{eq:etasingle}), we get that for all $l_1,l_2,i_1,i_2$
such that either $l_1\neq l$, $l_2\neq l$, or $i_1 \neq i_2$,
$$
 \bra{\eta_{l_1,i_1}}\left(I\otimes \ketbra{\widehat{t}}\right) \ket{\eta_{l_2,i_2}} = 0.
$$
Otherwise, $l_1=l_2=l$ and $i_1=i_2=i$ for some $i$ and we have
$$
 \bra{\eta_{l,i}}\left(I\otimes \ketbra{\widehat{t}}\right) \ket{\eta_{l,i}} = \frac{1}{L}.
$$
Hence we obtain
$$I \otimes \ketbra{\widehat{t}}|_{\calS_{prop1}}=\frac{1}{L} \sum_{i=0}^{2^N-1} \ketbra{\eta_{l,i}}$$
and similarly,
$$I \otimes \ketbra{\widehat{t \mns 1}}|_{\calS_{prop1}}=\frac{1}{L}\sum_{i=0}^{2^N-1} \ketbra{\eta_{l-1,i}}.$$
For the off-diagonal terms we see that
$$
 \bra{\eta_{l_1,i_1}}\left(\cphase\otimes \ket{\widehat{t}}\bra{\widehat{t \mns 1}}\right) \ket{\eta_{l_2,i_2}}=0
$$
if $l_1 \neq l$ or $l_2 \neq l - 1$. If $l_1=l$ and $l_2=l-1$ then using $\cphase=U_{lL}$, we get
$$\bra{\eta_{l,i_1}}\left(\cphase\otimes \ket{\widehat{t}}\bra{\widehat{t \mns 1}} \right) \ket{\eta_{l - 1,i_2}}
 = \frac{1}{L} \bra{i_1} \left( U_{lL} \cdots U_1 \right)^\dagger  \cphase U_{lL - 1} \cdots U_1 \ket{i_2} =
 \frac{1}{L} \langle i_1 | i_2  \rangle $$
which is $0$ if $i_1 \neq i_2$ and $\frac{1}{L}$ otherwise.
Hence $\cphase\otimes \ket{\widehat{t}}\bra{\widehat{t \mns 1}}|_{\calS_{prop1}}=
\frac{1}{L} \sum_{i=0}^{2^N-1} \ket{\eta_{l,i}}\bra{\eta_{l - 1,i}}$ and similarly for its Hermitian adjoint. This
establishes Eq. (\ref{eqn:hprop2}).

\paragraph{3. Restriction to $\calS_{prop}$:}

Let $\calS_{prop}$ be the $2^N$-dimensional space whose basis is given by the states
$$\ket{\eta_i} =
    \frac{1}{\sqrt{T+1}} \sum_{t=0}^T U_t\cdots U_1 \ket{i} \otimes \ket{\widehat t} = \frac{1}{\sqrt{T_2+1}} \sum_{l=0}^{T_2} \ket{\eta_{l,i}},$$
for $i \in \{0,\ldots,2^N-1\}$. Eq. (\ref{eqn:hprop2}) shows that $\calS_{prop}$ is an eigenspace of $H'$ of eigenvalue
$0$. Moreover, $H'$ is block-diagonal with $2^N$ blocks of size $T_2+1$. Each block is a matrix as in Eq. (\ref{eq:h}), multiplied by $1/L$.
As in Claim \ref{claim:gap} we see that the smallest non-zero eigenvalue of this Hamiltonian is $c/L T_2^2\geq c/T^2$ for some constant $c$. Now
we can apply Lemma \ref{lem:leak}. This time, we apply it inside $\calS_{prop1}$ with
\begin{align*}
 H_1&=\left(H_{out}+J_{in} H_{in}\right)|_{\calS_{prop1}}   &   H_2&=J_2 H'.
\end{align*}
Eigenvectors of $H_2$ orthogonal to $\calS_{prop}$ have eigenvalue at least $J=J_2 c/T^2$. As before, we can choose
$J_2=\poly(N)$ such that  $\lambda(H_1+H_2)$ is lower bounded by  $\lambda(H_{1}|_{\calS_{prop}})-\frac{1}{8}$. Hence,
in the remainder we consider
$$H_{out}|_{\calS_{prop}}+J_{in} H_{in}|_{\calS_{prop}}.$$

\paragraph{4. Restriction to $\calS_{in}$:}
The rest of the proof proceeds in the same way as in Section \ref{sec:previous}. Indeed, the subspace $\calS_{prop}$ is
isomorphic to the one in Section \ref{sec:previous} and both $H_{out}|_{\calS_{prop}}$ and $H_{in}|_{\calS_{prop}}$ are
the same Hamiltonians. So by another application of Lemma \ref{lem:leak} we get that the lowest eigenvalue of
$H_{out}|_{\calS_{prop}}+J_{in} H_{in}|_{\calS_{prop}}$ is lower bounded by
$\lambda(H_{out}|_{\calS_{in}})-\frac{1}{8}$. As in Section \ref{sec:previous}, we have
that $\lambda(H_{out}|_{\calS_{in}}) > 1-\eps$ if the circuit accepts with
probability less than $\eps$. Hence $\lambda(H)$, the lowest eigenvalue of the
original Hamiltonian $H$, is larger than $1-\eps - \frac{4}{8}=\frac{1}{2}-\eps$.

\end{proof}


\section{Perturbation Theory Proof}\label{sec:perturbation}

In this section we give an alternative proof of our main theorem. In Section \ref{ssec:perttheory},
we develop our perturbation theory technique. Since this technique might constitute a
useful tool in other Hamiltonian constructions, we keep the presentation as general as possible.
Then, in Section \ref{sec:gadget}, we present a specific application of our technique, the
three-qubit gadget.
Finally, in Section \ref{sec:32}, we use this gadget to complete the proof of the main theorem.

\subsection{Perturbation theory}\label{ssec:perttheory}

The goal in perturbation theory is to analyze the spectrum of the
sum of two Hamiltonians $\pert{H}=H+V$ in the case that $V$ has a small norm compared to the spectral gap of $H$.
One setting was described in the projection lemma. Specifically, assume $H$ has a zero
eigenvalue with the associated eigenspace $\calS$, whereas all other eigenvalues are greater than $\Delta\gg\|V\|$. The
projection lemma shows that in this case, the lowest eigenvalue of $\pert{H}$ is close to that of $V|_{\calS}$.
In this section we find a better approximation to $\Spec \pert{H}$ by considering certain correction terms that involve
higher powers of $V$. It turns out that these higher order correction terms include interesting interactions,
which will allow us to create an effective $3$-local Hamiltonian from $2$-local terms.
We remark that the projection lemma (for the entire lower part of the spectrum) can be obtained
by following the development done in this section up to the first order.

Before giving a more detailed description of the technique,
we need to introduce a certain amount of notation.
For two Hermitian operators $H$ and $V$, let $\pert{H}=H+V$. We refer to $H$ as the \emph{unperturbed Hamiltonian} and
to $V$ as the \emph{perturbation Hamiltonian}. Let $\lambda_{j}$, $|\psi_{j}\rangle$ be the eigenvalues and eigenvectors
of $H$, whereas the eigenvalues and eigenvectors of $\pert{H}$ are denoted by $\pert{\lambda}_{j}$,
$|\pert{\psi}_{j}\rangle$. In case of multiplicities, some eigenvalues might appear more than once. We order the
eigenvalues in a non-decreasing order
$$
  \lambda_{1}\le \lambda_{2} \le \dots\le\lambda_{\dim \calH},\qquad\quad
  \pert{\lambda}_{1}\le\pert{\lambda}_{2}\le \dots\le\pert{\lambda}_{\dim \calH}.
$$
In general, everything related to the perturbed Hamiltonian is marked with a tilde.

An important component in our proof is the \emph{resolvent} of $\pert{H}$, defined as
\begin{equation}\label{eq:Green}
\pert{G}(z)\,=\,\bigl(z I-\pert{H}\bigr)^{-1}\,=\, \sum_{j}\bigl(z -\pert{\lambda}_{j}\bigr)^{-1}
\bigl|\pert{\psi}_{j}\bigr\rangle\bigl\langle\pert{\psi}_{j}\bigr|.
\end{equation}
It is a meromorphic\footnote{A meromorphic function is analytic in all but a discrete subset of $\C$, and these
singularities must be poles and not essential singularities.} operator-valued function of the complex variable $z$ with
poles at $z=\pert{\lambda}_{j}$. In fact, for our purposes, it is sufficient to consider real $z$.\footnote{
The resolvent is the main tool in abstract spectral theory~\cite{Rudin}. In physics, it is known as the \emph{Green's
function}. Physicists actually use slightly different Green's functions that are suited for specific
problems.} Its usefulness comes from the fact that poles can be preserved under projections (while
eigenvalues are usually lost). Similarly, we define the resolvent of $H$ as $G(z)=(z I-H)^{-1}$.\footnote{
We can express $\pert{G}$ in terms of $G$ (where we omit the variable $z$):
$$
\pert{G}\,=\, \bigl(G^{-1}-V\bigr)^{-1} \,=\, G \bigl(I-VG\bigr)^{-1} \,=\, G+GVG+GVGVG+GVGVGVG+\cdots.
$$
This expansion of $\pert{G}$ in powers of
$V$ may be represented by Feynman diagrams~\cite{AGD}.
}

Let $\lambda_* \in \R$ be some cutoff on the spectrum of $H$.
\begin{definition}
Let $\calH=\calL_+ \oplus \calL_-$, where $\calL_+$ is the space spanned by eigenvectors of $H$ with eigenvalues
$\lambda \geq \lambda_*$ and $\calL_-$ is spanned by eigenvectors of $H$ of eigenvalue $\lambda<\lambda_*$. Let
$\Pi_\pm$ be the corresponding projection onto $\calL_\pm$. For an operator $X$ on $\calH$ define the operator
$X_{++}=X|_{\calL_+}=\Pi_+ X\Pi_+$ on $\calL_+$ and similarly $X_{--}=X|_{\calL_-}$. We also define $X_{+-}= \Pi_+ X
\Pi_-$ as an operator from $\calL_-$ to $\calL_+$, and similarly $X_{-+}$.
\end{definition}
With these definitions, in a representation of $\calH=\calL_+\oplus \calL_-$ both $H$ and $G$ are block diagonal and we
will omit one index for their blocks, i.e., $H_{+} \defeq H_{++}$, $G_+ \defeq G_{++}$ and so on. Note that
$G_{\pm}^{-1}=z I_{\pm}-H_{\pm}$. To summarize, we have:
\begin{eqnarray*}
&&\pert{H}=\left(\begin{array}{cc} \pert{H}_{++} &  \pert{H}_{+-}\\ \pert{H}_{-+} &  \pert{H}_{--} \end{array}\right) \quad V=\left(\begin{array}{cc} V_{++} &  {V}_{+-}\\ {V}_{-+} &  {V}_{--} \end{array}\right) \quad  H=\left(\begin{array}{cc} H_{+} &  0\\ 0 &  H_{-} \end{array}\right)\nonumber \\
&&\pert{G}=\left(\begin{array}{cc} \pert{G}_{++} &  \pert{G}_{+-}\\ \pert{G}_{-+} &  \pert{G}_{--} \end{array}\right)
\quad G=\left(\begin{array}{cc} G_{+} &  0\\ 0 &  G_{-} \end{array}\right)
\end{eqnarray*}
We similarly write $\calH=\pert{\calL}_+ \oplus \pert{\calL}_-$ according to the spectrum of $\pert{H}$ and the cutoff
$\lambda_*$. Finally, we define
\begin{equation*}
\Sigma_{-}(z)=z I_{-}- \pert{G}_{--}^{-1}(z).
\end{equation*}
This operator-valued function is called \emph{self-energy}.\footnote{As we will see later,
this defintion includes an $H_{-}$ term. This term is usually not considered part of
self-energy, but we have included it for notational convenience.}

With these notations in place, we can now give an overview of what follows.
Our goal is to approximate the spectrum of $\pert{H}|_{\pert{\calL}_-}$.
We will do this by showing that in some sense, the spectrum of $\Sigma_-(z)$ gives
such an approximation. To see why this arises,
notice that by definition of $\Sigma_-(z)$, we have
$ \pert{G}_{--}(z)=\bigl(z I_{-}-\Sigma_{-}(z)\bigr)^{-1}$.
In some sense, this equation is the analogue of Eq. \eqref{eq:Green} where $\Sigma_-(z)$ plays the role
of a Hamiltonian for the projected resolvent $\pert{G}_{--}(z)$.
However, $\Sigma_-(z)$ is in general $z$-dependent and not a fixed Hamiltonian.
Nonetheless, for certain choices of $H$ and $V$, $\Sigma_-(z)$ is
nearly constant in a certain range of $z$ so we can choose an {\em effective Hamiltonian} $H_{\eff}$
that approximates $\Sigma_-(z)$ in this range.
Our main theorem relates the spectrum of $H_{\eff}$ to that of $\pert{H}|_{\pert{\calL}_-}$.

\begin{theorem}\label{thm:perturbation}
Assume $H$ has a spectral gap $\Delta$ around the cutoff $\lambda_*$, i.e., all its eigenvalues are in $
(-\infty,\lambda_{-}]\cup[\lambda_{+},+\infty)$, where $\lambda_+=\lambda_* + \Delta/2$ and $\lambda_-=\lambda_* -
\Delta/2$. Assume moreover that $\|V\| < \Delta/2$. Let $\eps>0$ be arbitrary. Assume there exists an operator
$H_{\eff}$ such that $\Spec H_{\eff} \subseteq [c,d]$ for some $c<d<\lambda_*-\eps$ and moreover, the inequality
\begin{equation*}
\|\Sigma_{-}(z)-H_{\eff}\|\le\eps
\end{equation*}
holds for all $z\in[c-\eps,d+\eps]$. Then each eigenvalue $\pert{\lambda}_j$ of $\pert{H}|_{\pert{\calL}_-}$ is
$\eps$-close to the $j$th eigenvalue of $H_{\eff}$.
\end{theorem}

The usefulness of the theorem comes from the fact that $\Sigma_-(z)$ has a natural
series expansion, which can be truncated to obtain $H_{\eff}$.
This series may give rise to interesting terms; for example, in our application,
$2$-local terms in $H$ and $V$ lead to $3$-local terms in $H_\eff$.
To obtain this expansion, we start by expressing $\pert{G}$ in terms of $G$ as
\begin{equation*}
\pert{G}\,=\, \bigl(G^{-1}-V\bigr)^{-1} \,=\, \left(\begin{array}{cc} G_{+}^{-1}-V_{++} &  -V_{+-}\\ -V_{-+} &
G_{-}^{-1}-V_{--} \end{array}\right)^{-1}.
\end{equation*}
Then, using the block matrix identity
\[
\begin{pmatrix}
A & B\\ C & D
\end{pmatrix}^{-1} \,=\,
\left(\begin{array}{@{}cc@{}} \bigl(A-BD^{-1}C\bigr)^{-1} & -A^{-1}B\bigl(D-CA^{-1}B\bigr)^{-1}
\\[3pt]
-D^{-1}C\bigl(A-BD^{-1}C\bigr)^{-1} & \bigl(D-CA^{-1}B\bigr)^{-1}
\end{array}\right)
\]
we conclude that
\begin{equation*}
\pert{G}_{--}= \Bigl(G_{-}^{-1}-V_{--} -V_{-+}\bigl(G_{+}^{-1}-V_{++}\bigr)^{-1}V_{+-}\Bigr)^{-1}.
\end{equation*}
Finally, we can represent $\Sigma_{-}(z)$ using the series expansion $(I-X)^{-1}=I+X+X^2+\cdots$,
\begin{equation} \label{eq:self-energy}
\begin{aligned}
\Sigma_{-}(z)\, &=\,H_-+V_{--} \,+\, V_{-+}\bigl(G_{+}^{-1}-V_{++}\bigr)^{-1}V_{+-} \\[3pt]
  &=\,H_-+V_{--} \,+\, V_{-+}  G_{+} \bigl(I_+ -V_{++}G_{+} \bigr)^{-1}V_{+-} \\[3pt]
  &=\, H_- + V_{--} \,+\, V_{-+}G_{+}V_{+-} \,+\, V_{-+}G_{+}V_{++}G_{+}V_{+-} \,+\, V_{-+}G_{+}V_{++}G_{+}V_{++}G_{+}V_{+-} \,+\,\cdots.
\end{aligned}
\end{equation}

\begin{proof}[ of Theorem \ref{thm:perturbation}]
We start with an overview of the proof. We first notice
that, by definition, the eigenvalues of $\pert{H}|_{\pert{\calL}_-}$ appear
as poles in $\pert{G}$.
In Lemma \ref{lem1}, we show that these poles
also appear as poles of $\pert{G}_{--}$.
As mentioned before, this is the reason we work with resolvents.
In Lemmas \ref{lem2} and \ref{lem3} we relate these poles
to the eigenvalues of $\Sigma_-$ by showing that $z$ is a pole of $\pert{G}_{--}$
if and only if it is an eigenvalue of $\Sigma_-(z)$. In other words,
these are values of $z$ for which $\Sigma_-(z)$ has $z$ as an eigenvalue.
Finally, we complete the proof of the theorem by using the assumption
that $\Sigma_-(z)$ is close to $H_{\eff}$, so any eigenvalue of $\Sigma_-(z)$
must be close to an eigenvalue of $H_{\eff}$.
This situation is illustrated in Figure \ref{fig:evs}.

\begin{figure}[h]
\center{
 \epsfxsize=2.5in
     \epsfbox{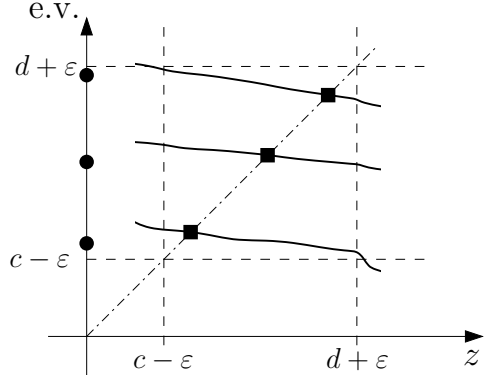}}
 \caption{The spectrum of $\Sigma_-(z)$ as a function of $z$ is indicated with solid curves. The boxes
 correspond to the spectrum of $\pert{H}|_{\pert{\calL}_-}$; they are those eigenvalues of $\Sigma_-(z)$ that lie
 on the dashed line $z=e.v.$ The dots indicate the spectrum of $H_{\eff}$, which approximates the spectrum of
 $\pert{H}|_{\pert{\calL}_-}$.
     }
 \label{fig:evs}
\end{figure}

We start with a simple lemma that says that if two Hamiltonians $H_1$, $H_2$
are close, their spectra must also be close.
It is a special case of Weyl's inequalities (see, e.g., Section III.2 in \cite{bhatia}).

\begin{lemma} \label{lem:spectrum}
Let $H_1,H_2$ be two Hamiltonians with eigenvalues $\mu_{1}\le \mu_{2} \le \ldots$ and $\sigma_{1} \le \sigma_{2} \le \ldots$.
Then, for all $j$, $|\mu_{j}-\sigma_{j}|\le\|H_1 - H_2\|$.
\end{lemma}

\begin{proof}
We will use a fact from the theory of Hermitian forms: if $X\le Y$ (i.e., if $Y-X$ is positive semidefinite),
then the operator $Y$ has at least as many positive and nonnegative eigenvalues as $X$. Let $\eps=\|H_1 - H_2\|$; then
\[
(\mu_{j}-\eps)I-H_2\le \mu_{j}I-H_1 \le(\mu_{j}+\eps)I-H_2.
\]
The operator $\mu_{j}I-H_1$ has at most $j-1$ positive and at least $j$ nonnegative eigenvalues. Hence
$(\mu_{j}-\eps)I-H_2$ has at most $j-1$ positive eigenvalues, and $(\mu_{j}+\eps)I-H_2$ has at least
$j$ nonnegative eigenvalues. It follows that $\sigma_{j} \in [\mu_{j}-\eps,\,\mu_{j}+\eps]$.
\end{proof}

The next lemma asserts that the poles of $\pert{G}_{--}$ in the range $(-\infty, \lambda_*)$ are in one-to-one
correspondence with the eigenvalues of $\pert{H}|_{\pert{\calL}_-}$. Hence we can recover the eigenvalues of
$\pert{H}|_{\pert{\calL}_-}$ from the poles of $\pert{G}_{--}$.
\begin{lemma}\label{lem1}
Let $\tilde{\lambda}$ be in $(-\infty, \lambda_*)$ and let $m \ge 0$ be its multiplicity as an eigenvalue of
$\pert{H}|_{\pert{\calL}_-}$. Then around $\tilde{\lambda}$, $\pert{G}_{--}$ is of the form
$(z-\tilde{\lambda})^{-1}A+O(1)$ where $A$ is a rank $m$ operator.
\end{lemma}
\begin{proof}
We first show that $\pert{\calL}_{-}\cap\calL_{+}=\{0\}$. Suppose the contrary, i.e., there is a nonzero vector
$|\xi\rangle\in\pert{\calL}_{-}\cap\calL_{+}$.  W.l.o.g.\ $\langle\xi|\xi\rangle=1$. Then we have
$\langle\xi|(H+V)|\xi\rangle\le \lambda_{*}$ (since $|\xi\rangle\in\pert{\calL}_{-}$) and $\langle\xi|H|\xi\rangle\ge
\lambda_{+}$ (since $|\xi\rangle\in\calL_{+}$). Hence $\langle\xi|V|\xi\rangle\le \lambda_{*}-\lambda_{+}=-\Delta/2$.
But this is impossible because $\|V\|<\Delta/2$.

Now, since $\pert{\calL}_{-}\cap\calL_{+}=\{0\}$, we have that $\Pi_-
\ket{\xi} \neq 0$ for all nonzero vectors $\ket{\xi} \in \pert{\calL}_-$.
 From Eq. (\ref{eq:Green}) we obtain
$$\pert{G}_{--}=\Pi_-\pert{G}\Pi_-=
    \sum_{j}(z-\pert{\lambda}_j)^{-1}\Pi_- |\pert{\psi}_{j}\rangle \langle\pert{\psi}_{j}| \Pi_-.$$
If the multiplicity of $\tilde{\lambda}$ is $m$ then the matrix $\sum |\pert{\psi}_{j}\rangle
\langle\pert{\psi}_{j}|$ of the corresponding eigenvectors has rank $m$. This implies that the matrix $\sum\Pi_-
\ket{\pert{\psi}_{j}} \bra{\pert{\psi}_{j}} \Pi_-$ also has rank $m$. Indeed, if there is some linear combination of
$\Pi_- \ket{\pert{\psi}_{j}}$ that sums to zero then taking the same linear combination of $\ket{\pert{\psi}_{j}}$ must
also sum to zero.
\end{proof}
The next two lemmas relate the spectrum of
$\pert{H}|_{\pert{\calL}_-}$ to the operator $\Sigma_{-}(z)$.
\begin{lemma}\label{lem2}
For any $z < \lambda_*$, the multiplicity of $z$ as an eigenvalue of $\pert{H}|_{\pert{\calL}_-}$ is equal to the
multiplicity of $z$ as an eigenvalue of $\Sigma_-(z)$.
\end{lemma}
\begin{proof}
Fix some $z < \lambda_*$ and let $m$ be its multiplicity as an eigenvalue of $\pert{H}$ (in particular, $m=0$ if $z$ is
not an eigenvalue of $\pert{H}$). In the neighborhood of $z$ the function $\pert{G}_{--}(w)$ has the form
\[
\pert{G}_{--}(w)= (w-z)^{-1}A+B+O\bigl(|w-z|\bigr),
\]
where by Lemma \ref{lem1}, $A$ is an operator of rank $m$.  We now consider $\pert{G}_{--}^{-1}(w)$. For
any $w < \lambda_{+}-\|V\|$ the norm of $G_+(w)$ is strictly less than $1/\|V\|$. Hence, by Eq. \eqref{eq:self-energy} we
see that all the poles of $\Sigma_-(w)$ lie on the interval $\bigl[\lambda_{+}-\|V\|,\,+\infty\bigr)$; in
particular $\pert{G}_{--}^{-1}(w) = w I_- - \Sigma_-(w)$ is analytic for $w\in(-\infty,\lambda_{*}]$. Hence we can write
\begin{equation*}
\pert{G}_{--}^{-1}(w)=wI_{-}-\Sigma_{-}(w) =C+D(w-z)+O\bigl(|w-z|^{2}\bigr).
\end{equation*}
We claim that the dimension of the null-space of $C$ is exactly $m$. Notice that this implies that $z$ is an $m$-fold
eigenvalue of $\Sigma_-(z)= zI_- - C$. By multiplying the two equations above, we obtain
\begin{equation*}
I_-=\pert{G}_{--}^{-1}(w)\pert{G}_{--}(w)= (w-z)^{-1}CA + (DA+CB) + O(|w-z|).
\end{equation*}
By equating coefficients, we obtain $CA=0$ and $DA+CB=I_-$. On one hand, $CA=0$ implies that the null-space of $C$ has
dimension at least $m$. On the other hand, the rank of $DA$ is at most ${\rm rank}(A)=m$. Since $I_-$ has full rank, the
dimension of the null-space of $CB$ must be at most $m$. This implies that the dimension of the null-space of $C$ must
also be at most $m$.
\end{proof}

We observe that the function $\Sigma_-(z)$ is monotone decreasing in the operator sense (i.e., if $z_1 \le z_2$ then
$\Sigma_-(z_1) - \Sigma_-(z_2)$ is positive semidefinite):
\begin{align*}
\frac{d\Sigma_{-}(z)}{dz}\,&=\, \frac{d}{dz}\Bigl(H_- + V_{--}+V_{-+}(z I_{+}-H_{+}-V_{++})^{-1}V_{+-}\Bigr) \\
 &=\, -V_{-+}(z I_{+}-H_{+}-V_{++})^{-2}V_{+-} \,\le\, 0.
\end{align*}
\begin{lemma}\label{lem3}
Let $\pert{\lambda}_j$ be the $j$th eigenvalue of $\pert{H}|_{\pert{\calL}_-}$. Then it is also the $j$th eigenvalue of
$\Sigma_-(\pert{\lambda}_j)$.
\end{lemma}
\begin{proof}
For any $z \in \R$, let $f_1(z)$ (resp., $f_2(z)$) be the number of eigenvalues not greater than $z$ of
$\pert{H}|_{\pert{\calL}_-}$ (resp., $\Sigma_-(z)$). When $z\rightarrow -\infty$, $f_1(z)$ is clearly 0. By the
monotonicity of $\Sigma_-$ we see that $f_2(z)$ is also 0. Using Lemma \ref{lem2} we see that as $z$ increases, both
numbers increase together by the same amount $m$ whenever $z$ hits an eigenvalue of $\pert{H}|_{\pert{\calL}_-}$ of
multiplicity $m$ (here we used again the monotonicity of $\Sigma_-$). Hence, for all $z$, $f_1(z)=f_2(z)$ and the lemma
is proven.
\end{proof}

We can now complete the proof of the theorem. By Lemma \ref{lem:spectrum} and our assumption on $H_{\eff}$, we have
that for any $z\in [c-\eps, d+\eps]$, $\Spec \Sigma_-(z)$ is contained in $[c-\eps, d+\eps]$. From this and the
monotonicity of $\Sigma_-$, we obtain that there is no $z \in (d+\eps,\lambda_*]$ that is an eigenvalue of
$\Sigma_-(z)$. Similarly, there is no $z<c-\eps$ that is an eigenvalue of $\Sigma_-(z)$. Hence, using Lemma \ref{lem2}
we see that $\Spec \pert{H}|_{\pert{\calL}_-}$ is contained in $[c-\eps, d+\eps]$. Now let $\pert{\lambda}_j \in
[c-\eps, d+\eps]$ be the $j$th eigenvalue of $\pert{H}|_{\pert{\calL}_-}$. By Lemma \ref{lem3} it is also the $j$th
eigenvalue of $\Sigma_-(\pert{\lambda}_j)$. By Lemma \ref{lem:spectrum} it is $\eps$-close to the $j$th eigenvalue of
$H_{\eff}$.

\end{proof}

\subsection{The Three-Qubit Gadget}\label{sec:gadget}

In this section we demonstrate how Theorem \ref{thm:perturbation} can be used to transform a 3-local Hamiltonian into a
2-local one. The complete reduction will be shown in the next section. From now we try to keep the discussion more
specialized to our $\QMA$ problem rather than presenting it in full generality as was done in Section
\ref{ssec:perttheory}.

Let $Y$ be some arbitrary 2-local Hamiltonian acting on a space $\calM$ of $N$ qubits. Also, let $B_1,B_2,B_3$ be
positive semidefinite Hamiltonians each acting on a different qubit (so they commute). We think of these four operators
as having constant norm. Assume we have the 3-local Hamiltonian
\begin{equation}\label{eq:3qgoal}
Y - 6 B_1 B_2 B_3.
\end{equation}
The factor 6 is added for convenience. Recall that in the {\locHam} problem we are interested in the lowest eigenvalue
of a Hamiltonian. Hence, our goal is to find a 2-local Hamiltonian whose lowest eigenvalue is very close to the lowest
eigenvalue of \eqref{eq:3qgoal}.

We start by adding three qubits to our system. For $j=1,2,3$, we denote the Pauli operators acting on the $j$th qubit
by $\sigma^{\alpha}_{j}$. Let $\delta >0$ be a sufficiently small constant. Our 2-local Hamiltonian is $\pert{H}=H+V$, where
\begin{align*}
  H&=-\frac{\delta^{-3}}{4} I\otimes \bigl(\sigma^{z}_{1}\sigma^{z}_{2} +\sigma^{z}_{1}\sigma^{z}_{3}
  +\sigma^{z}_{2}\sigma^{z}_{3}-3I\bigr)
\\[3pt]
  V&=X\otimes I- \delta^{-2}\bigl(B_{1}\otimes\sigma^{x}_{1} +B_{2}\otimes\sigma^{x}_{2} + B_{3}\otimes\sigma^{x}_{3}\bigr)
\\[3pt]
   X&=Y+\delta^{-1}(B_{1}^{2}+B_{2}^{2}+B_{3}^{2})
\end{align*}
The unperturbed Hamiltonian $H$ has eigenvalues $0$ and $\Delta \defeq \delta^{-3}$. Associated with the zero
eigenvalue is the subspace
$$
\calL_{-}=\calM\otimes\calC,\qquad \text{where}\quad \calC=\bigl(|000\rangle,|111\rangle\bigr).
$$
In the orthogonal subspace $\calC^\perp$ we have the states $\ket{001},\ket{010}$, etc. We may think of the subspace
$\calC$ as an effective qubit (as opposed to the three physical qubits); the corresponding Pauli operators are denoted
by $\sigma^{\alpha}_{\eff}$.

To obtain $H_\eff$, we now compute the self-energy $\Sigma_-(z)$ using the power expansion
in Eq. \eqref{eq:self-energy} up to the third order. There is no zeroth order term, i.e., $H_{-}=0$. For the remaining
terms, notice that $G_+ = (z-\Delta)^{-1}I_{\calL_+}$. Hence, we have
$$ \Sigma_{-}(z) = V_{--} + (z-\Delta)^{-1} V_{-+}V_{+-} + (z-\Delta)^{-2} V_{-+}V_{++}V_{+-} + (z-\Delta)^{-3} V_{-+}V_{++}V_{++}V_{+-} + \cdots.$$
The first term is $V_{--}=X\otimes I_{\calC}$ because a $\sigma^x$ term takes any state in $\calC$ to $\calC^\perp$. The expressions in the following terms are of the form
\begin{align*}
V_{-+}&=-\delta^{-2} \Bigl(B_1 \otimes \ket{000}\bra{100}+B_2 \otimes \ket{000}\bra{010}+B_3 \otimes \ket{000}\bra{001}+
\\[3pt]
&\qquad \qquad B_1 \otimes \ket{111}\bra{011}+B_2 \otimes \ket{111}\bra{101}+B_3 \otimes \ket{111}\bra{110}\Bigr)
\\[3pt]
V_{++}&=X \otimes I_{\calC^\perp}-\delta^{-2} \Bigl( B_1 \otimes \left(
\ket{001}\bra{101}+\ket{010}\bra{110}+\ket{101}\bra{001}+\ket{110}\bra{010}\right) +
\\[3pt]
&\qquad \qquad \qquad \qquad B_2 \otimes (\ldots )+ B_3 \otimes (\ldots ) \Bigr),
\end{align*}
where the dots denote similar terms for $B_2$ and $B_3$. Now, in the second term of $\Sigma_-(z)$, $V_{+-}$ flips one
of the physical qubits, and $V_{-+}$ must return it to its original state in order to return to the space $\calC$.
Hence we have $V_{-+}V_{+-} = \delta^{-4}(B_{1}^{2}+B_{2}^{2}+B_{3}^{2})\otimes I_{\calC}$. The third term is slightly
more involved. Here we have two possible processes. In the first process, $V_{+-}$ flips a qubit, $V_{++}$ acts with
$X\otimes I_{\calC^{\perp}}$, and finally $V_{-+}$ flips the qubit back. In the second process, $V_{+-}$, $V_{++}$, and
$V_{-+}$ flip all three qubits in succession. Thus,
\begin{equation} \label{eq:gadgetSigma}
\begin{aligned}
\hskip-5pt \Sigma_{-}(z)\, \hskip-20pt &\hskip20pt =\, X\otimes I_{\calC} +
(z-\Delta)^{-1}\delta^{-4}(B_{1}^{2}+B_{2}^{2}+B_{3}^{2})\otimes I_{\calC}
\\[3pt]
&+(z-\Delta)^{-2} \delta^{-4}(B_{1}XB_{1}+B_{2}XB_{2}+B_{3}XB_{3}) \otimes I_{\calC}
\\[3pt]
&-(z-\Delta)^{-2} \delta^{-6}\bigl(B_{3}B_{2}B_{1}+B_{2}B_{3}B_{1}+B_{3}B_{1}B_{2}+
B_{1}B_{3}B_{2}+B_{2}B_{1}B_{3}+B_{1}B_{2}B_{3}\bigr)\otimes \sigma^{x}_{\eff}
\\[3pt]
&+ O\bigl(\|V\|^{4}(z-\Delta)^{-3}\bigr).
\end{aligned}
\end{equation}
We now focus on the range $z=O(1) \ll \Delta$. In this range we have
$$(z-\Delta)^{-1}=-\frac{1}{\Delta}\Bigl(1-\frac{z}{\Delta}\Bigr)^{-1}=-\frac{1}{\Delta}+O(z/\Delta^2)=-\delta^3+O(\delta^6).$$
Simplifying, we obtain
$$
\Sigma_{-}(z) \,=\, \underbrace{Y\otimes I_{\calC} \,- \, 6 B_{1}B_{2}B_{3} \otimes \sigma^{x}_{\eff} }_{H_{\eff}}
    \,+\, O(\delta).
$$
Notice that $\|H_{\eff}\|=O(1)$ and hence we obtain that for all $z$ in, say, $[-2\|H_{\eff}\|, 2\|H_{\eff}\|]$ we
have
$$ \| \Sigma_-(z) - H_{\eff} \| = O(\delta).$$
We may now apply Theorem~\ref{thm:perturbation} with $c=-\|H_{\eff}\|$, $d=\|H_{\eff}\|$, and $\lambda_* = \Delta/2$ to
obtain the following result: Each eigenvalue $\pert{\lambda}_{j}$ from the lower part of $\Spec\pert{H}$ is
$O(\delta)$-close to the $j$-th eigenvalue of $H_{\eff}$. In fact, for our purposes, it is enough that the lowest
eigenvalue of $\pert{H}$ is $O(\delta)$-close to the lowest eigenvalue of $H_{\eff}$. It remains to notice that the
spectrum of $H_{\eff}$ consists of two parts that correspond to the effective spin states
$|+\rangle=\frac{1}{\sqrt{2}}\bigl(|0\rangle+|1\rangle\bigr)$ and
$|-\rangle=\frac{1}{\sqrt{2}}\bigl(|0\rangle-|1\rangle\bigr)$. Since $B_1 B_2 B_3$ is positive semidefinite, the
smallest eigenvalue is associated with $|+\rangle$. Hence, the lowest eigenvalue of $\pert{H}$ is equal to the lowest
eigenvalue of \eqref{eq:3qgoal}, as required.

\subsection{Reduction from 3-{\locHam} to 2-{\locHam}}\label{sec:32}

In this section we reduce the 3-{\locHam} problem to the 2-{\locHam} problem. By the $\QMA$-completeness of the 3-{\locHam} problem \cite{Kempe:03a}, this establishes Theorem \ref{thm:main_thm}.

\begin{theorem} \label{th:3to2}
There is a polynomial time reduction from the 3-{\locHam} problem to the 2-{\locHam} problem.
\end{theorem}
\begin{proof}
Recall that in the 3-{\locHam} problem (see Def. \ref{Def:localHam}) we are given
two constants $a$ and $b$ and a local Hamiltonian $H^{(3)}=\sum_{j} H_j$ such
that each $H_j$ is a 3-qubit term whose norm is at most $\poly(n)$.
Our goal in this proof is to transform $H^{(3)}$ into a 2-local Hamiltonian $H^{(2)}$
whose lowest eigenvalue is close to that of $H^{(3)}$.
We do this in two steps. The first is a somewhat technical step where we bring
$H^{(3)}$ into a convenient form. In the second step, we replace each 3-local term with
2-local terms by using the gadget construction of the previous section.
Before we continue with the proof, let us mention that it is crucial that
we apply the gadget construction to all 3-local terms {\em simultaneously}.
If instead we tried to apply the gadget construction sequentially,
we would end up with an exponential blowup in the norms (since each
application of the three-qubit gadget increases the norm by a multiplicative
factor).

\begin{lemma} \label{lem:mod3q}
The 3-local Hamiltonian $H^{(3)}$ can be represented as
\[H^{(3)} = c_r\left(Y\, - \, 6\sum_{m=1}^{M}B_{m1}B_{m2}B_{m3}\right)\]
where $Y$ is a 2-local Hamiltonian with $\|Y\|=O(1/n^6)$, $M=O(n^{3})$, each $B_{mi}$ is a one-qubit term of norm $O(1/n^3)$ that
satisfies $B_{mi}\ge \frac{1}{n^3} I $, and $c_r$ is a rescaling factor satisfying $1 \le c_r \le \poly(n)$.\footnote{
For the proof of Thm. \ref{th:3to2} we only need the property $B_{mi}\ge 0$. The stronger property $B_{mi}\ge \frac{1}{n^3} I $
will be used in Sec. \ref{sec:adiabatic}.}
\end{lemma}
\begin{proof}
First, we can assume without loss of generality that each $H_j$ acts on a different triple of qubits, and hence there are
at most $n^3$ such terms. Recall that any $3$-qubit Hermitian operator can be written as a linear combination with real
coefficients of the basis elements $\sigma^{\alpha}\otimes\sigma^{\beta}\otimes\sigma^{\gamma}$ where each of
$\sigma^\alpha,\sigma^\beta,\sigma^\gamma$
ranges over the four possible Pauli matrices $\{I,\sigma^x,\sigma^y,\sigma^z\}$.
Hence, for $M=O(n^3)$, we can write
\[
H^{(3)}= c_r\left(- 6 \sum_{m=1}^M c_m \cdot \sigma^{m,\alpha}\otimes\sigma^{m,\beta}\otimes\sigma^{m,\gamma}\right),
\]
where each $\sigma^{m,\alpha}$ is a Pauli matrix acting on one of the qubits,
and $c_r \le \poly(n)$ is chosen to be large enough so that $|c_m|\leq \frac{1}{n^9}$ for all $m=1,\ldots,M$.

We can now write
\begin{align*}
c_m \,\sigma^{m,\alpha}\otimes\sigma^{m,\beta}\otimes\sigma^{m,\gamma}&= \underbrace{\left(\frac{2}{n^3}I+
 n^6 c_m \sigma^{m,\alpha}\right)}_{B_{m1}}\otimes\underbrace{\left(\frac{2}{n^3}I+
 \frac{1}{n^3} \sigma^{m,\beta}\right)}_{B_{m2}}\otimes\underbrace{\left(\frac{2}{n^3}I+
 \frac{1}{n^3} \sigma^{m,\gamma}\right)}_{B_{m3}}\,+\,D_m
\end{align*}
where $D_m$ is $2$-local. Since
$|c_m| \leq 1/n^9$ we have that $B_{mi} \ge \frac{1}{n^3}I$ and  $\|D_m\|=O(1/n^9)$.
\end{proof}
We now replace each term $-6B_{m1}B_{m2}B_{m3}$ by a three-qubit gadget. More specifically, let $\delta$ be
a sufficiently small inverse polynomial in $n$ to be chosen later. We consider the Hamiltonian $H^{(2)}=c_r \pert{H}$, $\pert{H}=H+V$,
acting on a system of $n+3M$ qubits, where
\begin{align}
H\,&=\,-\,\frac{\delta^{-3}}{4} \sum_{m=1}^{M} I\otimes \bigl(\sigma^{z}_{m1}\sigma^{z}_{m2}
+\sigma^{z}_{m1}\sigma^{z}_{m3} +\sigma^{z}_{m2}\sigma^{z}_{m3}-3I\bigr),
\nonumber \\[5pt]
V\,&= Y\otimes I+\,{} \delta^{-1}\sum_{m=1}^{M}(B_{m1}^{2}+B_{m2}^{2}+B_{m3}^{2})\otimes I
\nonumber \\
 & \qquad \qquad  - \delta^{-2} \sum_{m=1}^{M} \bigl(B_{m1}\otimes\sigma^{x}_{m1} + B_{m2}\otimes\sigma^{x}_{m2} +
B_{m3}\otimes\sigma^{x}_{m3}\bigr). \label{eq:multi_gadget_hamiltonian}
\end{align}
As before, let $\Delta = \delta^{-3}$ be the spectral gap of $H$. Notice that the spectrum of $H$ includes not only $0$
and $\Delta$, but also $2\Delta, 3 \Delta, \ldots , M \Delta$. Associated with the zero eigenvalue is the subspace
spanned by all the zero-subspaces of the gadgets. Using $\|B_{mi}\| \leq O(1/n^3)$ and $M=O(n^3)$ we get
$\|V\|=O(\delta^{-2})<\Delta/2$.

The calculation of $\Sigma_{-}$ is quite similar to the one-gadget case (cf.\ Eq.~(\ref{eq:gadgetSigma})). Each gadget
contributes an independent term. Terms up to the third order can only include processes that involve one gadget.
Indeed, in order to involve two gadgets, one has to flip a qubit from one gadget and from another gadget, and then flip
both qubits back. Moreover, since only one gadget is involved, $G_+$ can be replaced by $(z-\Delta)^{-1}I_{\calL_+}$ as
before. From the fourth order onwards, processes start to include cross-terms between different gadgets. However, we
claim that their contribution is only $O(\delta)$, as long as $|z|=O(1)$. Indeed, in this range, the
eigenvalues of $G_+$, which are $(z-\Delta)^{-1}$, $(z-2\Delta)^{-1}$, $\ldots$, are all at most $O(\delta^3)$ in absolute value
while the norm of each of the $V$ terms is at most $O(\delta^{-2})$. To summarize, for $|z|=O(1)$,
\begin{equation}\label{eq:biggadget}
\Sigma_{-}(z) \,=\, \underbrace{Y\otimes I_{\calC} \, - \,6\sum_{m=1}^{M} B_{m1}B_{m2}B_{m3} \otimes
\bigl(\sigma^{x}_{m}\bigr)_{\eff}}_{H_{\eff}} \,+\,
 O(\delta).
\end{equation}
Since $\|H_\eff\| \le O(1)$, we can apply Theorem \ref{thm:perturbation} with $c=-\|H_\eff\|$, $d=\|H_\eff\|$ and $\lambda_*=\Delta/2$. We
obtain that the smallest eigenvalue of $\pert{H}$ is $O(\delta)$-close to that of $H_{\eff}$.
The spectrum of $H_{\eff}$ consists of $2^M$ parts, corresponding to subspaces spanned
by setting each effective spin state to either $|+\rangle$ or $|-\rangle$.
Since $B_{m1}B_{m2}B_{m3}\ge 0$, the smallest eigenvalue of $H_{\eff}$ is achieved in the subspace
where all effective spin states are in the $\ket{+}$ state.
In this subspace, $H_{\eff}$ is identical to $H^{(3)}/c_r$. Hence, the smallest eigenvalue of
$H^{(2)}=c_r \pert{H}$ is $O(c_r \delta)$-close to that of $H^{(3)}$.
We complete the proof by choosing $\delta=c'/c_r$ for some small enough constant $c'$.
\end{proof}

\section{2-local Universal Adiabatic Computation}\label{sec:adiabatic}

In this section we show that adiabatic computation with 2-local Hamiltonians is equivalent to ``standard''
quantum computation in the circuit model. In order to prove such an equivalence, one has to show that each model can
simulate the other. One direction is already known: it is not too hard to show that any polynomial time adiabatic
computation can be efficiently simulated by a quantum circuit \cite{farhiad}. Hence, it remains to show that adiabatic
computation with 2-local Hamiltonians can efficiently simulate any quantum circuit.
In \cite{Aharonov:04a} it is shown that adiabatic computation with $3$-local Hamiltonians can
efficiently simulate any quantum circuit. We obtain our result by combining their result with the
techniques in our second proof.

Let us briefly mention the main ideas behind adiabatic computation. For more details see \cite{Aharonov:04a} and
references therein. In adiabatic computation, we consider a time-dependent Hamiltonian $H(s)$ for $s \in [0,1]$ acting
on a quantum system. We initialize the system in the groundstate of the initial Hamiltonian $H(0)$. This groundstate
is required to be some simple quantum state that is easy to create. We then slowly modify the Hamiltonian from
$s=0$ to $s=1$. We say that the adiabatic computation is {\em successful} if the final state of the system
is close to the groundstate of
$H(1)$. The adiabatic theorem (see, e.g., \cite{ben,AmbainisR04}) says that
if the Hamiltonian is modified slowly enough, the adiabatic computation is successful.
In other words, it gives an upper bound on the running time of an adiabatic computation.
For our purposes, it is enough to know that this bound is polynomial if
for any $s\in [0,1]$, the norm of $H(s)$, as well as that of its first and second derivatives,
is bounded by a polynomial, and the spectral gap of $H(s)$ is larger than some inverse polynomial.

In \cite{Aharonov:04a} it is shown how to transform an arbitrary quantum circuit
into an efficient $3$-local adiabatic computation. To establish this, they define
a $3$-local time-dependent Hamiltonian $H^{(3)}(s)$
with the following properties. First, the Hamiltonian acts on a system of $n$ qubits, where $n$
is some constant times the number of gates in the circuit.
Second, the groundstate of $H^{(3)}(0)$ is very easy to create (namely, it is the all zero state),
and the groundstate of $H^{(3)}(1)$ is some state that encodes the result of
the quantum circuit. Third, for all $s\in [0,1]$, the spectral gap of $H^{(3)}(s)$
is bounded from below by an inverse polynomial in $n$ and the norm
of $H^{(3)}(s)$, as well as that of its first and second derivatives, is bounded by
some polynomial in $n$. Together with the adiabatic theorem, these properties imply that adiabatic
computation according to $H^{(3)}(s)$ is efficient.
Finally, let us mention that $H^{(3)}(s)$, as defined in \cite{Aharonov:04a}, is linear in $s$, that is,
$H^{(3)}(s)=(1-s)H^{(3)}(0)+sH^{(3)}(1)$. This property will be useful in our proof.

The following is the main theorem of this section.
\begin{theorem}\label{th:adiabatic}
Any quantum computation can be efficiently simulated by an adiabatic computation with 2-local Hamiltonians.
\end{theorem}
\begin{proof}
Given a quantum circuit, let $H^{(3)}(s)$ be the time-dependent Hamiltonian
of \cite{Aharonov:04a} as described above.
The idea of the proof is to apply the gadget construction of Sec. \ref{sec:32}
to $H^{(3)}(s)$ for any $s \in [0,1]$, thereby creating
a 2-local time-dependent Hamiltonian $H^{(2)}(s)$. Some care needs to be taken to ensure
that the resulting time-dependent Hamiltonian is smooth enough as a function
of $s$. We therefore describe how this is done in more detail.

We start by writing $H^{(3)}(s)$ in a form similar to that given
by Lemma \ref{lem:mod3q}. Since $H^{(3)}(s)$ is linear in $s$, we can write
$$
H^{(3)}(s)= c_r\left(- 6 \sum_{m=1}^M c_m(s) \cdot \sigma^{m,\alpha}\otimes\sigma^{m,\beta}\otimes\sigma^{m,\gamma}\right),
$$
where $M=O(n^3)$, each $c_m(s)$ is a linear function of $s$, and $c_r \le \poly(n)$ is chosen to be large enough
so that $|c_m(s)| \le \frac{1}{n^9}$ for all $m$ and all $s\in [0,1]$. Notice that $c_r$ is a fixed
scaling factor, used for all $s\in [0,1]$.
Following the proof of Lemma \ref{lem:mod3q}, we write
\begin{equation*}
H^{(3)}(s) = c_r\left(Y(s)\, - \, 6\sum_{m=1}^{M}B_{m1}(s) B_{m2} B_{m3} \right)
\end{equation*}
where by our construction, $Y(s)$ and $B_{m1}(s)$ are linear in $s$, whereas $B_{m2}$
and $B_{m3}$ are independent of $s$. Finally, we define $H^{(2)}(s)=c_r \tilde{H}(s)$,
where $\tilde{H}(s)=H+V(s)$ and the Hamiltonians $H$ and $V(s)$ are defined as
in Eq. \eqref{eq:multi_gadget_hamiltonian}. The parameter $\delta$ will
be chosen later to be some small enough inverse polynomial in $n$.

In the rest of the proof, we show that adiabatic computation according to $H^{(2)}(s)$ can
be used to simulate the given quantum circuit. We start by proving two lemmas
that, together with the adiabatic theorem, imply that the running time of the
adiabatic computation is polynomial in $n$.


\begin{lemma}\label{lem:adlem2}
For any $s\in [0,1]$, $\|H^{(2)}(s)\|$, $\|\frac{d}{ds}H^{(2)}(s)\|$, and $\|\frac{d^2}{ds^2}H^{(2)}(s)\|$
are upper bounded by a polynomial in $n$.
\end{lemma}
\begin{proof}
Recall that $Y(s)$ and $B_{m1}(s)$ are linear in $s$. Together with the definition of $H^{(2)}$,
this implies that $H^{(2)}(s)$ is a degree two polynomial in $s$, i.e., we can write
$H^{(2)}(s)=A+sB+s^2C$ for some Hermitian matrices $A,B,C$. It is not hard to see
that the norm of each of these matrices is bounded by some polynomial in $n$.
This implies that the norm of $H^{(2)}(s)$, of its first derivative $B+2sC$, and of
its second derivative $2C$ are bounded by some polynomial in $n$.
\end{proof}

\begin{lemma}\label{lem:adlem1}
For any $s \in [0,1]$, the spectral gap of $H^{(2)}(s)$ is lower bounded by an inverse polynomial in $n$.
\end{lemma}
\begin{proof}
As shown in Sec. \ref{sec:32}, the lower part of the spectrum of $H^{(2)}(s)$ is
$O(c_r \delta)$-close to the spectrum of $c_r H_\eff(s)$.
Hence, by choosing $\delta$ to be a small enough inverse polynomial
in $n$, we see that it is enough to show that the spectral gap
of $c_r H_\eff(s)$ is at least some inverse polynomial in $n$.

The spectrum of $c_r H_\eff(s)$ consists
of $2^M$ parts, corresponding to all possible settings for the effective
qubits. The part corresponding to the subspace in which all effective qubits
are in the $\ket{+}$ state is identical to the spectrum of
$H^{(3)}(s)$. Hence, we know that in this subspace the spectral gap
is at least some inverse polynomial in $n$. We now claim that the
lowest eigenvalue in all other $2^M-1$ subspaces is greater
than that in the all $\ket{+}$ subspace by at least some
inverse polynomial in $n$. Indeed, the restriction
of $c_r H_\eff(s)$ to any such subspace is given
by $H^{(3)}(s)$ plus a nonzero number of terms of
the form $12 c_r B_{m1}(s)B_{m2}B_{m3}$. The claim follows
from the fact that $B_{m1}(s)B_{m2}B_{m3} \geq \frac{1}{n^9} I$.
\end{proof}

To complete the proof, we need to argue about the groundstate of $H^{(2)}(0)$ and that of $H^{(2)}(1)$.
To this end, we use the following lemma, which essentially says that if $H_\eff$ has a spectral gap, then Theorem
\ref{thm:perturbation} not only implies closeness in spectra but also in the groundstates.

\begin{lemma}\label{lem:adlem3}
Assume that $H,V,H_\eff$ satisfy the conditions of Theorem \ref{thm:perturbation} with some $\eps>0$.
Let $\lambda_{\eff,i}$ denote the $i$th eigenvalue of $H_\eff$ and $\ket{\pert{v}}$ (resp., $\ket{v_\eff}$) denote
the groundstate of $\pert{H}$ (resp., $H_\eff$).
Then, under the assumption $\lambda_{\eff,2}>\lambda_{\eff,1}$,
$$ | \langle \pert{v} | v_\eff \rangle | \ge 1 - \frac{2\|V\|^2}{(\lambda_+ - \lambda_{\eff,1} - \eps)^2} - \frac{4\eps}{\lambda_{\eff,2}-\lambda_{\eff,1}}.$$
\end{lemma}

Before we prove the lemma, let us complete the proof of the theorem.
Recall that in our case $\eps=O(\delta)$, $\|V\|=O(\delta^{-2})$, $\lambda_+ = \delta^{-3}$, $|\lambda_{\eff,1}| \le O(1)$
and $\lambda_{\eff,2}-\lambda_{\eff,1}=1/\poly(n)$. Hence, the first error term in the above bound is $O(\delta^2)$
while the second is $O(\delta\cdot \poly(n))$. Therefore, by choosing $\delta$ to be a
small enough inverse polynomial in $n$, we can guarantee that the groundstate of $H^{(2)}(s)$
is close to the groundstate of $H_\eff(s)$.
In particular, the groundstate of $H^{(2)}(1)$, which is
the output of the adiabatic computation, is
close to the groundstate of $H_\eff(1)$. The latter is $\ket{v_1} \otimes \ket{+}^{\otimes M}$,
where $\ket{v_1}$ is the groundstate of $H^{(3)}(1)$.
By simply tracing out the $3M$ gadget qubits, we can recover
$\ket{v_1}$ from this groundstate, and therefore obtain
the output of the quantum circuit. Similarly,
the groundstate of $H^{(2)}(0)$, which is the state to which the
system should be initialized, is close to the groundstate
of $H_\eff(0)$. The latter is $\ket{v_0} \otimes \ket{+}^{\otimes M}$,
where $\ket{v_0}$ is the groundstate of $H^{(3)}(0)$.
We therefore initialize the system by setting the original $n$ qubits
to $\ket{v_0}$ and the $M$ gadgets to the effective $\ket{+}$ state.
This state is close to the groundstate of $H^{(2)}(0)$, and
since the adiabatic computation is unitary, this approximation does not
affect the output by much.

It remains to prove the lemma.
\begin{proof}[ of Lemma \ref{lem:adlem3}]
Let $\ket{\pert{v}_-}=\Pi_-\ket{\pert{v}}/\|\Pi_-\ket{\pert{v}}\|$ be the normalized projection of $\ket{\pert{v}}$ on
the space $\calL_-$. We first show that $\ket{\pert{v}_-}$ is close to $\ket{\pert{v}}$. By Theorem
\ref{thm:perturbation}, we know that $\pert{\lambda}_1 \le \lambda_{\eff,1}+\eps$. Hence,
$$ \| \Pi_+ \pert{H} \ket{\pert{v}}\| = \pert{\lambda}_1 \| \Pi_+ \ket{\pert{v}}\| \le (\lambda_{\eff,1}+\eps) \|\Pi_+ \ket{\pert{v}}\| $$
and
$$ \| \Pi_+ \pert{H} \ket{\pert{v}} \| = \| \Pi_+ H \ket{\pert{v}} + \Pi_+ V \ket{\pert{v}} \|
   \ge  \|\Pi_+ H \ket{\pert{v}}\| - \|V\| \ge \lambda_+ \| \Pi_+ \ket{\pert{v}}\| - \|V\|.$$
By combining the two inequalities we obtain
$$ \| \Pi_+ \ket{\pert{v}}\| \le \frac{\|V\|}{\lambda_+ - \lambda_{\eff,1} - \eps},$$
from which we see that
$$ \alpha \defeq |\langle \pert{v} | \pert{v}_- \rangle|
   = \|\Pi_- \ket{\pert{v}}\| \ge \|\Pi_- \ket{\pert{v}}\|^2 \ge 1 - \frac{\|V\|^2}{(\lambda_+ - \lambda_{\eff,1} - \eps)^2}. $$

Our next step is to show that $\ket{\pert{v}_-}$ is close to $\ket{v_\eff}$. For this we need to consider the proof of
Theorem \ref{thm:perturbation}. We start by taking Lemma \ref{lem1} with $\pert{\lambda} = \pert{\lambda}_1$. The lemma
says that $A$ is a matrix of rank $1$. By looking at the proof, it is easy to see that $A$ is in fact
$\Pi_-\ketbra{\pert{v}}\Pi_-$. Next, Lemma \ref{lem2} implies that $\pert{\lambda}_1$ is an eigenvalue of multiplicity
$1$ of $\Sigma_-(\pert{\lambda}_1)$. In fact, from the proof it follows that the corresponding eigenvector is exactly
$\Pi_-\ket{\pert{v}}$ (since the null space of $C$ is equal to the span of $A$). By normalizing, this is exactly
$\ket{\pert{v}_-}$. But by our assumption, $\|\Sigma_{-}(z)-H_{\eff}\|\le\eps$ for all $z\in [c-\eps,d+\eps]$ and in
particular
\begin{equation*}
\|\Sigma_{-}(\pert{\lambda}_1)-H_{\eff}\|\le\eps.
\end{equation*}
 From this we obtain that
$$ \bigl|\bra{\pert{v}_-} (\Sigma_{-}(\pert{\lambda}_1)-H_{\eff}) \ket{\pert{v}_-} \bigr| \le \eps $$
and hence
$$ \bra{\pert{v}_-} H_{\eff} \ket{\pert{v}_-} \le \pert{\lambda}_1+\eps \le \lambda_{\eff,1} + 2\eps$$
where we again used that $\pert{\lambda}_1 \le \lambda_{\eff,1} + \eps$. Since $H_{\eff}$ has a spectral gap, this
indicates that $\ket{\pert{v}_-}$ must be close to $\ket{v_{\eff}}$. Indeed, let $\beta=|\langle \pert{v}_- | v_\eff
\rangle|$. Then,
$$ \bra{\pert{v}_-} H_{\eff} \ket{\pert{v}_-} \ge \beta^2 \lambda_{\eff,1} + (1-\beta^2)\lambda_{\eff,2} =
    \lambda_{\eff,1} + (1-\beta^2) (\lambda_{\eff,2}-\lambda_{\eff,1}).$$
By combining the two inequalities we obtain
$$ 1-\beta^2 \le \frac{2\eps}{\lambda_{\eff,2}-\lambda_{\eff,1}}.$$
Summarizing,
\begin{align*}
|\langle \pert{v} | v_\eff \rangle| &=
   |\langle \pert{v} | \pert{v}_- \rangle \langle \pert{v}_- | v_\eff \rangle +
   \langle \pert{v} | ( I - \ketbra{\pert{v}_-}) | v_\eff \rangle |\\
   & \ge \alpha \cdot \beta - \sqrt{(1-\alpha^2)(1-\beta^2)}
   \ge \alpha \cdot \beta - \frac{1}{2}\bigl((1-\alpha^2)+ (1-\beta^2)\bigr) \\
   & \ge \bigl( 1- (1-\alpha) - (1-\beta) \bigl) - \bigl((1-\alpha) + (1-\beta)\bigr) = 1 - 2(1-\alpha) - 2(1-\beta) \\
   & \ge 1 - \frac{2\|V\|^2}{(\lambda_+ - \lambda_{\eff,1} - \eps)^2} - \frac{4\eps}{\lambda_{\eff,2}-\lambda_{\eff,1}}.
\end{align*}
\end{proof}
\end{proof}

\section{Conclusion}\label{sec:conclude}

Some interesting open questions remain. First, perturbation theory has allowed us to perform the first reduction {\em inside} $\QMA$. What other problems can be solved using this technique? Second, there exists an intriguing class between $\NP$ (in fact, $\MA$) and
$\QMA$ known as $\QCMA$. It is the class of problems that can be verified by a quantum verifier with a {\em classical}
proof. Can one show a separation between $\QCMA$ and $\QMA$? or perhaps show they are equal? Third, Kitaev's original
5-local proof has the following desirable property. For any {\Yes} instance produced by the reduction there exists a
state such that each individual 5-local term is very close to its groundstate. Note that this is a stronger property
than the one required in the {\locHam} problem. Using a slight modification of Kitaev's original construction, one can
show a reduction to the 4-{\locHam} problem that has the same property. However, we do not know if this property can be
achieved for the 3-local or the 2-local problem.

\section*{Acknowledgments} 
Discussions with Sergey Bravyi and Frank Verstraete are gratefully
acknowledged. JK is supported by ACI S\'ecurit\'e Informatique,
2003-n24, projet ``R\'eseaux Quantiques", ACI-CR 2002-40 and EU 5th
framework program RESQ IST-2001-37559, and by DARPA and Air Force
Laboratory, Air Force Materiel Command, USAF, under agreement number
F30602-01-2-0524, and by DARPA and the Office of Naval Research under
grant number FDN-00014-01-1-0826 and during a visit supported in part by the National
Science Foundation under grant EIA-0086038 through the Institute for
Quantum Information at the California Institute of Technology. AK
is supported in part by the National Science Foundation under
grant EIA-0086038. OR is supported by an Alon
Fellowship, the Binational Science Foundation, the Israel Science Foundation,
and the Army Research Office grant DAAD19-03-1-0082.
Part of this work was carried out during a visit of OR at LRI, Universit\'e de
Paris-Sud and he thanks his hosts for their hospitality and
acknowledges partial support by ACI S\'ecurit\'e Informatique,
2003-n24, projet ``R\'eseaux Quantiques".

\bibliographystyle{alpha}
\newcommand{\etalchar}[1]{$^{#1}$}

\end{document}